\pdfoutput=1 %arXiv admin added this line
\documentclass[a4paper, 12pt]{article}
\usepackage[ansinew]{inputenc}
\usepackage{amssymb}
\usepackage{amsfonts}
\usepackage{amsmath}
\usepackage{hyperref}
\usepackage{dsfont} % for symbols like the identity: \mathds{1}
\usepackage{graphicx} % for graphics

\numberwithin{equation}{section} % Section-Nummer vor Formelnummer
\newcommand{\captionfonts}{\small}
\makeatletter  % Allow the use of @ in command names
\long\def\@makecaption#1#2{%
  \vskip\abovecaptionskip
  \sbox\@tempboxa{{\captionfonts #1: #2}}%
  \ifdim \wd\@tempboxa >\hsize
    {\captionfonts #1: #2\par}
  \else
    \hbox to\hsize{\hfil\box\@tempboxa\hfil}%
  \fi
  \vskip\belowcaptionskip}
\makeatother   % Cancel the effect of \makeatletter
\begin{document}

\begin{titlepage}
\begin{center}
\vspace*{10mm}
\textsc{\huge{Quantum Entanglement \\ and Geometry}} \\ \vspace{30mm}
Diplomarbeit \\ zur Erlangung des akademischen Grades \\ ,,Magister der Naturwissenschaften'' \\ an der \\ \textsc{\large{Universität Wien}} \\ \vspace{25mm}
eingereicht von \\ \large{Andreas Gabriel} \\ \vspace{20mm}
\normalsize betreut von \\ \large{Ao. Univ. Prof. Dr. Reinhold A. Bertlmann} \\ \vspace{45mm}
Wien, Juni 2009
\end{center}
\end{titlepage}

\tableofcontents \newpage

\vspace*{10mm}
\section{Introduction}
Entanglement is one of the most nonclassical phenomenae in quantum physics, in the sense of being responsible for many effects that strongly contradict the very foundations of classical physics (such as local realism).\\ There are many different definitions of entanglement, some of which are very mathematical, others being rather close to the experiment and practical aplication. Simply speaking, two (or more) particles are entangled, if each of them cannot be fully described without the other, so that the combined system contains more information than the individual systems do.\\
\\
The concept of quantum entanglement was first perceived by Erwin Schrödin\-ger (who called it 'Verschränkung' in German, which only later was translated to 'Entanglement') in the early years of quantum mechanics in 1935\cite{schroed}. Soon Einstein, Podolsky and Rosen (EPR) discovered some of the extraordinary properties that followed from this strange new feature (so extraordinary in fact, that they believed quantum mechanics to be 'incomplete', rather than to consider these consequences to be real\cite{epr}). However, for a very long time both the works by Schrödinger and EPR were not taken seriously.\\
Almost 30 years later, in 1964, John Bell conceived a gedankenexperiment\cite{bell} which should turn out to be the means to settle this discussion once and for all, proving EPR's doubts to be wrong and making way for a huge new field of research in doing so.\\
Soon more and more highly interesting, sometimes intriguingly contraintuitive and often technologically very promising applications of entanglement, in particular the field of quantum information, were discovered.\\
Today entanglement and quantum information theory are very popular and steadily advancing subjects. While the first practical applications (such as quantum cryptography\cite{crypto}) are already about to make their way into everyday use, some areas (such as multipartite entanglement) of this field are merely beginning to be understood.\\
\\
When studying entanglement and related topics, one soon finds that geometry plays a very important role in many of these and often allows a highly intuitive and vivid understanding.\\
The aim of this work is to give a structured basic overview over the field of quantum entanglement and quantum information theory, concentrating mainly on this geometric viewpoint and on bipartite systems.\\
After giving an introduction into the mathematical formalism of quantum information theory in section 2, the very nature of entanglement will be discussed (3), followed by a detailed explanation of the most common and useful means to detect (4) and quantify (5) entanglement (with focus on the geometrical aspects). Finally, these tools will be used to thoroughly analyse the Hilbert spaces of quantum states, paying special attention to two of the most important classes of systems in quantum information theory (6), namely systems of two QuBits ($2\otimes2$ dimensions, e.g. two spin-½-particles) and systems of two QuTrits ($3\otimes3$ dimensions, e.g. two spin-1-particles).\\
\newpage

\section{Formalism and Basics}
\subsection{Hilbert Spaces and State Vectors}
\subsubsection{Bits, Dits, Qubits and Qudits}
In classical information theory, one usually deals with bits (i.e. variables that can assume the values 0 or 1) and -- less often -- also with higher dimensional generalisations thereof, dits (which can assume the values 0, 1, ..., $d-2$ and $d-1$). In quantum mechanics, superpositions of different states are also physically realised, which turns out to offer whole new possibilities, thus making quantum information theory a completely new field that can only remotely be compared to classical information theory.\\
The quantum analogy to a bit -- a quantum bit or qubit -- is a (normalised) complex superposition of the values 0 and 1. Hence, unlike a state of a classical bit (which can only have two different values), a state $\left|\Psi\right\rangle$ of a qubit can assume infinitely many different values of the form
\begin{equation} \left|\Psi\right\rangle = a \left|0\right\rangle + b \left|1\right\rangle \end{equation}
with the normalisation $\left|a\right|^2 + \left|b\right|^2 = 1$. Since the states $\left|0\right\rangle$ and $\left|1\right\rangle$ need to be orthogonal, the underlying Hilbert space is $\mathcal{H} = \mathds{C}^2$.\\
The classical generalisation of a bit to a dit also has a quantum analogue, the quantum dit or qudit, which can consequently assume all values that are superpositions of $d$ different states and lives on the Hilbert space $\mathcal{H} = \mathds{C}^d$. Apart from the qubit, especially the qutrit ($d=3$) also plays a very important role in quantum information theory..\\

\subsubsection{State Vectors}
Since the state of a qubit is described by a vector $\left|\Psi\right\rangle \in \mathds{C}^2$, it can be realised for example as the spin components of a spin-½ particle or the polarisation components of a photon. Therefore, the basis $\{\left|0\right\rangle, \left|1\right\rangle\}$ is also often denoted by $\{\left|\uparrow\right\rangle, \left|\downarrow\right\rangle\}$ or $\{\left|H\right\rangle, \left|V\right\rangle\}$ (where the latter stands for 'horizontal' and 'vertical' polarisation).\\
For convenience reasons, the basis vectors are usually chosen such that $\left|0\right\rangle$ is the first standard unit-vector and $\left|1\right\rangle$ is the second (in the case of qudits, this can easily be generalised to $\left|n\right\rangle$ being the $(n+1)$st standard unit vector). This choice of basis is called the computational basis.\\
The scalar product of two state vectors $\left|\Psi\right\rangle$ and $\left|\Phi\right\rangle$ is defined as
\begin{equation} \left\langle\Phi|\Psi\right\rangle = \left(\left|\Phi\right\rangle\right)^{*}\cdot\left|\Psi\right\rangle \end{equation}
where '$\cdot$' is the standard scalar product and the asterisk denotes complex conjugation.\\

\subsubsection{Product Spaces}
In quantum information theory, it is of central interest to describe several particles with a single state. These states are elements of a product space $\mathcal{H} = \mathcal{H}^{A} \otimes \mathcal{H}^{B} \otimes ... \otimes \mathcal{H}^{X}$, where $\mathcal{H}^{\alpha}$ are the state spaces of the individual systems, respectively. This work concentrates on bipartite systems, therefore the considered Hilbert spaces will be of the form $\mathcal{H} = \mathcal{H}^{A} \otimes \mathcal{H}^{B}$.\\
The dimension of such product spaces is
\begin{equation} \mathrm{dim}\mathcal{H} = \mathrm{dim}\mathcal{H}^{A} \mathrm{dim}\mathcal{H}^{B} = d_{1} d_{2} \end{equation}
where $d_{1,2}$ are the dimensions of the subspaces $\mathcal{H}^{A, B}$. In short notation, the space $\mathcal{H} = \mathds{C}^{d_{1}} \otimes \mathds{C}^{d_{2}}$ is also often referred to as $d_{1}\otimes d_{2}$.\\
Clearly, each pair of unipartite states $\{\left|\Psi^{A}\right\rangle, \left|\Psi^{B}\right\rangle\}$ corresponds to a bipartite state $\left|\Psi^{AB}\right\rangle = \left|\Psi^{A}\right\rangle\otimes\left|\Psi^{B}\right\rangle$. However, there are elements of the product space, which are not products of elements of the respective subspaces (but linear combinations of such). The individual parts of such states can obviously not be described by individual, mutually independant state vectors. Such states are called entangled and will be the main topic of this work.\\
Bases of product spaces are induced by bases of the subspaces. These are usually denoted by
\begin{equation} \left|i,j\right\rangle = \left|i\right\rangle \otimes \left|j\right\rangle \end{equation}
where $\left|i\right\rangle$ is a basis vector in the first subspace and $\left|j\right\rangle$ is a basis vector in the second subspace.\\

\subsection{Operators}
Operators acting on a Hilbert space $\mathcal{H}$ are represented by $d\times d$-dimensional matrices (where $d = \mathrm{dim}\mathcal{H}$). All operators on $\mathcal{H}$ form the Hilbert-Schmidt space $\mathcal{B}$, which is a Hilbert space itself (and therefore also often denoted by $\mathcal{H}$). It is equipped with the scalar product
\begin{equation} \left\langle A | B\right\rangle = \mathrm{Tr}\left(A^{\dagger}\cdot B\right) \end{equation}
which induces the norm
\begin{equation} \left\|A\right\| = \sqrt{\left\langle A | A\right\rangle} \label{hsnorm}\end{equation}
However, there also are other relevant norms on the Hilbert-Schmidt space. The n-norm is defined by
\begin{equation} \left\|A\right\| _{n} = \sqrt[n]{\mathrm{Tr}\left((A^{\dagger}A)^{\frac{n}{2}}\right)} \end{equation}
where the dagger $^{\dagger}$ denotes hermitean conjugation.\\
Most importantly, the 1-norm $\left\|.\right\|_{1}$ is also called the trace (class) norm, which obviously equals the sum of the absolute eigenvalues, and the 2-norm $\left\|.\right\|_{2}$ is equal to the standard Hilbert-Schmidt norm (\ref{hsnorm}).\\
An operator $A$ is said to be positive semidefinite ($A \geq 0$), if all of its eigenvalues are greater than or equal to zero. For simplicity, these operators are often simply called positive operators.\\

\subsection{Density Matrices}
\subsubsection{Definition}
In order to use the vector state formalism, complete knowledge of a quantum state is required. If for example the phase of an investigated state was unknown, the remaining information would be useless, as a state with 'averaged' phase would vanish. This problem can be solved by using the density matrix formalism, which includes the whole vector state formalism and also offers additional features.\\
The density matrix $\rho$ of a state $\left|\Psi\right\rangle$ is the operator defined as the outer product
\begin{equation} \rho = \left|\Psi\right\rangle\left\langle\Psi\right| \label{puredm}\end{equation}
This case, in which full information about the state is at hand and the density matrix assumes the above form, is called a pure state.\\
If however a state is not completely known, but only the probabilities $\{p_{i}\}$, with which it is one out of several states $\{\left|\Psi_{i}\right\rangle\}$ is known, it is called a mixed state and the density matrix assumes the form
\begin{equation} \rho = \sum_{i} p_{i} \left|\Psi_{i}\right\rangle\left\langle\Psi_{i}\right| \end{equation}
where evidently $p_{i} \geq 0$ and $\sum_{i} p_{i} = 1$.\\
All density matrices have the following properties:
\begin{itemize}
	\item Hermiticity: $\rho = \rho^{\dagger}$
	\item Normalisation: $\mathrm{Tr}(\rho) = 1$
	\item Positivity: $\rho \geq 0$
\end{itemize}
Furthermore, for pure states it follows from the definition (\ref{puredm}) that $\rho^{2} = \rho$ and thus $\mathrm{Tr}(\rho^{2}) = 1$, while for mixed states $\mathrm{Tr}(\rho^{2}) < 1$. In fact, the quantity
\begin{equation} M=\frac{d}{d-1}\left(1-\mathrm{Tr}(\rho^{2})\right) \end{equation}
can be considered a measure for the mixedness of a state (where $d = \mathrm{dim}\mathcal{H}$). It is known as the linear entropy and assumes its minimal value of zero for all pure states and its maximal value of one for the state $\omega = 1/d \ \mathds{1}$, which evidently is the maximally mixed state and is also referred to as the trace state.\\
It also follows from the definition (\ref{puredm}), that density matrices of pure states are projectors onto the corresponding state vector, which is their eigenvector to the eigenvalue one, while all other eigenvalues vanish.\\
Since density matrices are a more general class of objects that vector states (in the sense, that all vector states can be written as density matrices, but not vice versa), most of the time density matrices are used in quantum information theory. For convenience reasons, the word 'state' will refer to density matrices throughout this work, in contrast to vector states.\\

\subsubsection{Correlations}
Multipartite systems can be described by density matrices very much in the same way it can be described by vector states, that is, if two systems are described by the density matrices $\rho^{A}$ and $\rho^{B}$, the state of the composite system is 
\begin{equation}\label{proddm} \rho^{AB} = \rho^{A}\otimes\rho^{B} \end{equation}
Since this work concentrates on bipartite systems, the superscripts for these states will mostly be omitted -- it will always be clear from the context, what kind of state is meant.\\
Like in the case of vector states, also the Hilbert-Schmidt product space $\mathcal{B}=\mathcal{B}^{A}\otimes\mathcal{B}^{B}$ contains states, that cannot be expressed as products like in (\ref{proddm}). In the case of density matrices however, this can have two reasons. Firstly, analogously to the vector states, a state can be entangled, i.e. consist of entangled vector states (this will be discussed in further detail throughout this work). Secondly, the subsystems of a state can be classically correlated by mixing of product states. Consider for example the state
\begin{equation} \rho = \frac{1}{2}\left(\left|0,0\right\rangle\left\langle 0,0\right|+\left|1,1\right\rangle\left\langle 1,1\right|\right) \end{equation}
which is a convex sum of two product states. It is correlated in the sense that any measurement in the basis $\left|0\right\rangle, \left|1\right\rangle$ in either subsystem will always yield the same result. This is a purely classical effect and must not be confused with entanglement (i.e. quantum correlations).\\

\subsubsection{Reduced Density Matrices}
Evidently, the opposite operation to combining two Hilbert spaces to a product space is to reduce a composite state to one of its subsystems, discarding the other one. This is achived by tracing over the discarded subsystem, thus averaging over all correlations (if there are any) and being left with a unipartite state that in general contains less information about the corresponding subsystem than the composite state did, described by the reduced density matrices
\begin{equation}\begin{split} \rho^{A} = \mathrm{Tr}_{B}\left(\rho\right) \\
\rho^{B} = \mathrm{Tr}_{A}\left(\rho\right) \end{split}\end{equation}
where $\mathrm{Tr}_{X}$ is the partial trace over the subsystem $X$.\\
In the case of product states, this operation is fully reversible, i.e. the composite state can always be recovered by combining the reduced density matrices
\begin{equation} \rho = \rho^{A}\otimes\rho^{B} \end{equation}
while for other states, the tensor product
\begin{equation} \rho^{AB} = \rho^{A}\otimes\rho^{B} \end{equation}
will yield a different state $\rho^{AB} \neq \rho$, which contains less information than the original one, since all correlations were traced out and lost.\\

\subsubsection{Decompositions\label{decomp}}
The definition of mixed density matrices
\begin{equation} \rho = \sum_{i} p_{i} \left|\Psi_{i}\right\rangle\left\langle \Psi_{i}\right| \end{equation}
is not bijective in the sense that there is no unique decomposition of $\rho$ into an ensemble $(\{p_{i}\}, \{\left|\Psi_{i}\right\rangle)\}$, except for pure states. In most cases, there are in fact infinitely many such decompositions, such that it makes only little sense to speak of a mixed state consisting of a certain set of pure states, but only that a mixed state can be represented by such an ensemble.\\
Apart from decompositions into pure states, density matrices can of course also be decomposed mathematically into any basis of the underlying Hilbert-Schmidt space. Since this concept finds many applications in quantum information theory, the most important of these bases shall be presented here\cite{bloch}.\\

\begin{center}\emph{Pauli matrices}\end{center}
The most simple such basis consists of the Pauli matrices, which together with the identity matrix form a basis for all $2\times2$-dimensional matrices (or all hermitean $2\times2$-dimensional matrices, if all coefficients are held real). The three Pauli matrices are defined as
\begin{equation}\label{pauli} \sigma_{1} = \begin{pmatrix} 0 & 1 \\ 1 & 0 \end{pmatrix} \ , \ \sigma_{2} = \begin{pmatrix} 0 & -i \\ i & 0 \end{pmatrix} \ , \ \sigma_{3} = \begin{pmatrix} 1 & 0 \\ 0 & -1 \end{pmatrix} \end{equation}
Apart from their function as a basis, the Pauli matrices are also the observables corresponding to spin-measurements in the x-, y- and z-direction, respectively, for spin-½ particles and are also the generators of the SU(2).\\

\begin{center}\emph{Gell-Mann matrices}\end{center}
Originally, the Gell-Mann matrices (GMMs) were introduced as a basis (together with the identity matrix) for $3\times3$-dimensional hermitean matrices, however, later they were generalised to arbitrary dimensions.\\
For $d \times d$ dimensional matrices, there are $d(d-1)/2$ symmetric GMMs 
\begin{equation} \label{ggm} \lambda_{s}^{j,k} = \left|j\right\rangle\left\langle k\right| + \left|k\right\rangle\left\langle j\right| \end{equation}
as well as $d(d-1)/2$ antisymmetric GMMs 
\begin{equation} \lambda_{a}^{j,k} = -i \left|j\right\rangle\left\langle k\right| + i \left|k\right\rangle\left\langle j\right| \end{equation}
and $(d-1)$ diagonal GMMs
\begin{equation} \lambda_{d}^{l} = \sqrt{\frac{2}{l(l+1)}}\left(\sum_{m=1}^{l}\left|m\right\rangle\left\langle m\right| - l\left|l+1\right\rangle\left\langle l+1\right|\right) \end{equation}
with $1 \leq j < k \leq d$ and $1 \leq l \leq d-1$.\\
For $d=2$, the GMMs are equal to the Pauli matrices. For $d=3$, they read
\newpage
\[ \lambda_{s}^{1,2} = \begin{pmatrix} 0 & 1 & 0 \\ 1 & 0 & 0 \\ 0 & 0 & 0 \end{pmatrix} \ , \ \lambda_{s}^{1,3} = \begin{pmatrix} 0 & 0 & 1 \\ 0 & 0 & 0 \\ 1 & 0 & 0 \end{pmatrix} \ , \ \lambda_{s}^{2,3} = \begin{pmatrix} 0 & 0 & 0 \\ 0 & 0 & 1 \\ 0 & 1 & 0 \end{pmatrix}, \]
\begin{equation} \lambda_{a}^{1,2} = \begin{pmatrix} 0 & -i & 0 \\ i & 0 & 0 \\ 0 & 0 & 0 \end{pmatrix} \ , \ \lambda_{a}^{1,3} = \begin{pmatrix} 0 & 0 & -i \\ 0 & 0 & 0 \\ i & 0 & 0 \end{pmatrix} \ , \ \lambda_{a}^{2,3} = \begin{pmatrix} 0 & 0 & 0 \\ 0 & 0 & -i \\ 0 & i & 0 \end{pmatrix}, \end{equation}
\[ \lambda_{d}^{1} = \begin{pmatrix} 1 & 0 & 0 \\ 0 & -1 & 0 \\ 0 & 0 & 0 \end{pmatrix} \ , \ \lambda_{d}^{2} = \frac{1}{\sqrt{3}} \begin{pmatrix} 1 & 0 & 0 \\ 0 & 1 & 0 \\ 0 & 0 & -2 \end{pmatrix} \]
The generalisations to higher dimensions are straightforward.\\
Since all GMMs are hermitean themselves, all coefficients in a decomposition are real.\\

\begin{center}\emph{Weyl operators}\end{center}
Another basis for $d \times d$-dimensional matrices that has proven to be quite useful in quantum information theory is the Weyl operator basis, which consists of $d^{2}$ unitary and mutually orthogonal matrices $W_{m,n}$, defined as
\begin{equation}\label{weyl} W_{m,n} = \sum_{k=0}^{d-1} e^{\frac{2 \pi i}{d} k n} \left|k\right\rangle\left\langle k+m\right| \end{equation}
where $0 \leq m,n \leq d-1$ and $(k+m)$ is to be understood modulo $d$. Note that $U_{0,0} = \mathds{1}$.\\
For $d=2$, the four Weyl operators correspond to the identity and the three Pauli matrices: $U_{0,0} = \mathds{1}, \ U_{0,1} = \sigma_{1}, \ U_{1,0} = \sigma_{3}, \ U_{1,1} = i \sigma_{2}$.\\
For $d=3$, the Weyl operators assume the form
\newpage
\[ W_{0,1} = \begin{pmatrix} 0 & 1 & 0 \\ 0 & 0 & 1 \\ 1 & 0 & 0 \end{pmatrix}, \ W_{0,2} = \begin{pmatrix} 0 & 0 & 1 \\ 1 & 0 & 0 \\ 0 & 1 & 0 \end{pmatrix}, \]
\[ W_{1,0} = \begin{pmatrix} 1 & 0 & 0 \\ 0 & e^{\frac{2 \pi i}{3}} & 0 \\ 0 & 0 & e^{-\frac{2 \pi i}{3}} \end{pmatrix}, \ W_{1,1} = \begin{pmatrix} 0 & 1 & 0 \\ 0 & 0 & e^{\frac{2 \pi i}{3}} \\ e^{-\frac{2 \pi i}{3}} & 0 & 0 \end{pmatrix}, \ W_{1,2} = \begin{pmatrix} 0 & 0 & 1 \\ e^{\frac{2 \pi i}{3}} & 0 & 0 \\ 0 & e^{-\frac{2 \pi i}{3}} & 0 \end{pmatrix}, \]
\begin{equation} W_{2,0} = \begin{pmatrix} 1 & 0 & 0 \\ 0 & e^{-\frac{2 \pi i}{3}} & 0 \\ 0 & 0 & e^{\frac{2 \pi i}{3}} \end{pmatrix}, \ W_{2,1} = \begin{pmatrix} 0 & 1 & 0 \\ 0 & 0 & e^{-\frac{2 \pi i}{3}} \\ e^{\frac{2 \pi i}{3}} & 0 & 0 \end{pmatrix}, \ W_{2,2} = \begin{pmatrix} 0 & 0 & 1 \\ e^{-\frac{2 \pi i}{3}} & 0 & 0 \\ 0 & e^{\frac{2 \pi i}{3}} & 0 \end{pmatrix} \end{equation} \\

\begin{center}\emph{The Schmidt decomposition}\end{center}
Apart from decompositions for density matrices, there is also an important decomposition for composite vector states, known as the Schmidt de\-com\-posit\-ion\cite{schmidt}. It decomposes a bipartite state vector $\left|\Psi\right\rangle$ into a product basis, minimising the number of terms, i.e.
\begin{equation} \left|\Psi\right\rangle = \sum_{i=1}^{k} \left|\phi^{A}_{i}\right\rangle \otimes \left|\phi^{B}_{i}\right\rangle \end{equation}
The minimised number of terms $k$ is known as the Schmidt-rank of $\left|\Psi\right\rangle$. It is easy to see that $1 \leq k \leq d_{min}$, where $d_{min} = \min(d_{1}, d_{2})$ is the lower of the two subsystem's dimensions.\\
The Schmidt rank of a state also equals the rank of its reduced density matrices.\\
\newpage

\section{Entanglement and Distillation}
\subsection{Entanglement}
There are various ways in which entanglement can be seen and characterised\cite{bruss} -- for example it would usually be described very differently by an experimental and a mathematical physicist. In this section, it shall primarily be seen as a resource for performing various tasks, such as quantum computation\cite{compute} or quantum cryprography\cite{crypto}, thus avoiding philosophical and mathematical difficulties and concentrating on the applicational point of view (providing motivation for dealing with quantum information theory in the first place and in particular the motivation for this work).\\
In all these applications, quantum states are needed to be exchanged between two or more parties (usually refered to as Alice, Bob, Charlie, etc.). In general however, one does not have the means to transmit these states perfectly loss-free -- quantum channels are mostly noisy and quantum states are rather fragile. In classical communications, error correction protocols enable faithful communication even through imperfect channels by sending multiple copies of the data. Since it is impossible to copy a quantum state (as stated in the no-cloning-theorem\cite{nocloning}), this is not an option in quantum communication. Instead, it appears to be a good way to use quantum teleportation\cite{teleport} (thus, not sending the particle through a quantum channel, but using entanglement aided by a classical communications channel to transmit it), to send the state without losses. The obvious problem here is, that faithful teleportation requires maximally entangled pure states to be shared between the acting parties. Which ever of the parties creates these states, as soon as one of the entangled particles is sent to another party, the state will become mixed and less entangled -- again, due to noisy quantum channels and interaction with the environment.\\
So, under certain conditions, it is rather easy for Alice and Bob to get any number of nonmaximally entangled mixed states (since they can create and share these states arbitrarily many times), where they actually need maximally entangled ones. Hence, there is need for a means to 'distill' these mixed states back to maximally entangled pure states.\\

\subsection{Distillation}
The problem was solved by Bennet et al.\cite{distill}, who thought of a way for Alice and Bob to increase both entanglement and purity at cost of the number of their shared states. Since this first work, many others have been published in this field, providing several different protocols for distilling entangled states. Still, all follow the same basic way:
\begin{enumerate}
	\item Alice and Bob share a number of nonmaximally entangled states
	\item They both perform local measurements on their particles
	\item They tell each other the outcomes of these measurements (via a classical channel)
	\item Depending on these outcomes being equal or odd, they either discard the measured pair of particles or keep them. In the latter case, the particles are now more strongly entangled than they were before.
	\item These steps can be repeated as often as necessary, increasing the entanglement per pair of particles each time.
\end{enumerate}
Although the optimal protocol for distilling a state (i.e. the protocol requiring the least input states per maximally entangled output state) in general may depend on the state itself, an upper bound for the efficiency of all such protocols can be given\cite{bbpssw}, since -- considering entanglement as a resource -- the total entanglement in any system can not increase (under the given circumstances, i.e. only allowing Alice and Bob to perform local operations). Hence, if (in an appropriate measure, which will be discussed in section 5) the combined entanglement of all input states for any protocol equals the entanglement of one maximally entangled output state, this protocol is maximally efficient and cannot be exceeded by any other distillation protocol.\\

\begin{figure}[ht!]\centering\includegraphics[width=80mm]{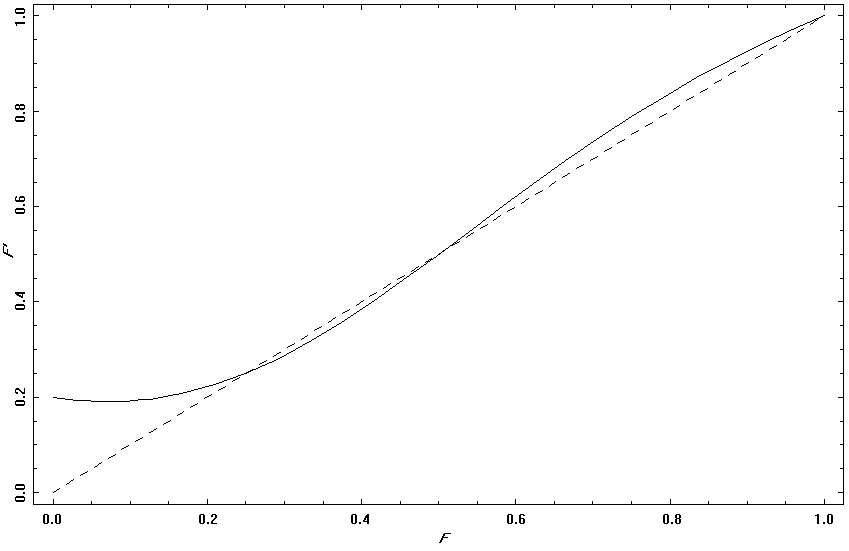}\caption[Efficiency of the BBPSSW-protocol]{Efficiency of the BBPSSW-protocol. The solid line shows the purity F' after applying the protocol, whereas the dashed line represents the purity F before.}\label{fig:dist_fid}\end{figure}

\begin{center}\emph{The BBPSSW-Protocol}\end{center}
The best known distillation protocol (and the only one that shall be discussed in this work) is the so called BBPSSW-protocol\cite{bbpssw}, which is designed for distillation of two-qubit-states (nevertheless, it can be generalised to higher dimensions). It works as follows:
\begin{enumerate}
	\item Alice and Bob share n pairs of mixed entangled states $\rho$ ($\rho^{\otimes n}$)
	\item They first apply random unitary rotations to all pairs
	\begin{equation} \rho \mapsto U_{rot} \ \rho \ U_{rot}^{\dagger} \end{equation}
	thus transforming these states into rotationary invariant states of the form
	\begin{equation}\begin{split}	W_{F} = F & \left|\Psi^{-}\right\rangle\left\langle \Psi^{-}\right| \ + \ \frac{1-F}{3}\left|\Psi^{+}\right\rangle\left\langle \Psi^{+}\right| \ +\\
	 + \ \frac{1-F}{3} & \left|\Phi^{-}\right\rangle\left\langle \Phi^{-}\right| \ + \ \frac{1-F}{3}\left|\Phi^{+}\right\rangle\left\langle \Phi^{+}\right| \end{split}\end{equation}
	for some F, where
	\begin{equation}\begin{split}\label{bellstates}	\left|\Psi^{\pm}\right\rangle = \frac{1}{\sqrt{2}}(\left|01\right\rangle \pm \left|10\right\rangle) \\
	\left|\Phi^{\pm}\right\rangle = \frac{1}{\sqrt{2}}(\left|00\right\rangle \pm \left|11\right\rangle)\end{split}\end{equation}
	are the Bell states, which are maximally entangled.\\
	The state W$_{F}$ is called a Werner state\cite{werner} of purity F and -- since only local unitary operations were performed -- is equivalent to the original state (thus in particular containing the same amount of entanglement).
	\item Alice and Bob now pick two pairs on which to perform further transformations in the following way:
	\begin{enumerate}
		\item Alice performs rotations by $\pi$ rad around the y-axis on both her particles, converting the $\left|\Psi^{-}\right\rangle$ fraction into $\left|\Phi^{+}\right\rangle$ and vice versa.
		\begin{equation} W_{F} \mapsto W'_{F} \end{equation}
		where
		\begin{equation}\begin{split}\label{werner} W'_{F} = F & \left|\Phi^{+}\right\rangle\left\langle \Phi^{+}\right| \ + \ \frac{1-F}{3}\left|\Psi^{+}\right\rangle\left\langle \Psi^{+}\right| \ +\\
	 + \ \frac{1-F}{3} & \left|\Phi^{-}\right\rangle\left\langle \Phi^{-}\right| \ + \ \frac{1-F}{3}\left|\Psi^{-}\right\rangle\left\langle \Psi^{-}\right| \end{split}\end{equation}
		\item Alice and Bob apply a biliteral XOR (or BXOR\cite{qugates}) on these two pairs.
		\begin{equation} W'_{F} \mapsto (U_{XOR}\otimes U_{XOR}) \ W'_{F} \ (U_{XOR}\otimes U_{XOR})^{\dagger} \end{equation}
		with the XOR-quantum-gate\cite{bbpssw}
		\begin{equation}\begin{split}U_{XOR} := & \left|\uparrow_{source}\uparrow_{target}\right\rangle\left\langle\uparrow_{source}\downarrow_{target}\right| + \left|\uparrow_{source}\downarrow_{target}\right\rangle\left\langle\uparrow_{source}\uparrow_{target}\right| + \\ + & \left|\downarrow_{source}\downarrow_{target}\right\rangle\left\langle\downarrow_{source}\downarrow_{target}\right| + \left|\downarrow_{source}\uparrow_{target}\right\rangle\left\langle\downarrow_{source}\uparrow_{target}\right| \end{split}\end{equation}
		\item Both perform spin measurements along the z-axis on the pair that was used as target in the previous step. Note that this step is the only nonunitary (i.e. irreversible) action performed in the protocol.
		\item Alice and Bob communicate their measurement results to each other classically. If they match, the source pair is kept (otherwise it is discarded).
		\item Finally, Alice may perform another rotation around the y-axis in order to transform the resulting state back into a Werner state.
		\end{enumerate}		
	\item The resulting state now has the new purity
	\begin{equation} F' = \frac{F^2 + \frac{1}{9}(1-F)^2}{F^2+\frac{2}{3}F(1-F)+\frac{5}{9}(1-F)^2} \end{equation}
	which is satisfies $F' > F$ for all $F >$ ½ (see Fig. \ref{fig:dist_fid}).
	\item Step 3 can be repeated arbitrarily often, increasing the purity (i.e. the entanglement per pair) each time.
\end{enumerate}
Note that although the efficiency of this process, i.e. the number of distilled maximally entangled states per less entangled input state, tends to zero, there are methods to achieve nonzero distillation rates by more complicated protocols, allowing to distill any number of maximally entangled states from a finite number of input states\cite{bbpssw}.\\
The question remains, whether mixed entangled states can always be distilled to maximally entangled pure states and if not so, how these undistillable ('bound') entangled states can be recognised.\\

\subsection{Bound Entanglement\label{be}}
Questions about the existence and identification of bound entangled states are rather complicated, since there are infinitely many possible distillation protocols and there is no reason why a state that cannot be distilled by one of them should not be distillable by any other one.\\
Fortunately, a general necessary and sufficient criterion for distillability of a state can be given via its partial transposition (i.e. a transposition in one of its subsystems, while the other subsystem is left unchanged). The partial transposition of a state $\rho$ is defined by\cite{dclb}
\begin{equation} \rho^{T_{A}} = \sum^{d_{A}}_{i,j=1} \sum^{d_{B}}_{k,l=1} \left\langle i,k\right|\rho\left| j,l\right\rangle \ \left|j,k\right\rangle\left\langle i,l\right| \end{equation}
where $d_{A}$ and $d_{B}$ are the dimensions of the subspaces A and B, respectively.\\
A state $\rho$ is distillable iff there is an integer $n$ such that the inequality
\begin{equation} \left\langle\Psi\right|(\rho^{T_{A}})^{\otimes n}\left|\Psi\right\rangle \geq 0 \end{equation}
(where $\rho^{\otimes n} = \rho \otimes \rho ... \otimes \rho$ is the n-fold copy of $\rho$) is violated by a Schmidt rank 2 vector $\left|\Psi\right\rangle$\cite{horo_be}.\\
From this follows the much more simple (although weaker) condition, that $\rho^{T_{A}}$ must be a nonpositive operator if $\rho$ is free entangled (i.e. distillable). Such a state $\rho$ is called a NPT-state (nonpositive partial transposition), while otherwise it is called a PPT-state (positive partial transposition).\\
Note that since the eigenvalues of a matrix do not depend on the choice of basis, neither does the positivity (while the eigenvectors do). Also it does not matter, which of the subsystems is being transposed, since an overall transposition does not change the eigenvalues.\\
Using this, one just has to find an entangled PPT-state (which as well is not a trivial task, but nevertheless is possible) in order to prove the existence of bound entanglement. The first state proven to be bound entangled was the two-qutrit state\cite{horo_be}
\begin{equation} \rho_{a} = \frac{1}{8a+1} 
\begin{pmatrix}
	a & 0 & 0 & 0 & a & 0 & 0 & 0 & a \\
	0 & a & 0 & 0 & 0 & 0 & 0 & 0 & 0 \\
	0 & 0 & a & 0 & 0 & 0 & 0 & 0 & 0 \\
	0 & 0 & 0 & a & 0 & 0 & 0 & 0 & 0 \\
	a & 0 & 0 & 0 & a & 0 & 0 & 0 & a \\
	0 & 0 & 0 & 0 & 0 & a & 0 & 0 & 0 \\
	0 & 0 & 0 & 0 & 0 & 0 & \frac{1+a}{2} & 0 & \frac{\sqrt{1-a^2}}{2} \\
	0 & 0 & 0 & 0 & 0 & 0 & 0 & a & 0 \\
	a & 0 & 0 & 0 & a & 0 & \frac{\sqrt{1-a^2}}{2} & 0 & \frac{1+a}{2}
\end{pmatrix}\end{equation}
with $0 < a < 1$. However, many more examples followed, making sure that bound entangled states are not at all rare (although hard to detect, as will be seen in the next section) and do not form a set of measure zero on the underlying Hilbert space.\\
An even more difficile task is to prove the existence or nonexistence of NPT bound entanglement. Since the nonpositivity of the partially transposed density matrix is not a sufficient criterion for distillability, this question cannot be trivially answered and has not been definitely solved yet. There exists evidence however, that strongly suggests the existence of NPT bound entanglement\cite{nptbound1, nptbe2}.\\
It seems quite obvious, that bound entanglement is not very useful, since it is a rather weak kind of entanglement (as mentioned previously, all Werner states with purity F $>$ ½ can be distilled, hence, all bound entangled states have to be states of lower purity -- similar bounds exist for higher dimensional generalisations of the discussed Werner state) and cannot be distilled to higher purity. Surprisingly, even this weak form of entanglement suffices for various quantum-informational tasks\cite{becrypto, bebell}. Also, there still seems to be a chance to ''quasi-distill`` bound entangled states\cite{activatebe}.\\

\begin{center}\emph{Quasi-Distillation of Bound Entanglement\label{activateboundent}}\end{center}
If Alice and Bob share only a small number of (or even only one) nonmaximally free entangled states but a large pool of bound entangled states, there is a possibility for them to transfer some of the entanglement from these bound entangled states into the free entangled ones. The main difference between this procedure and genuine distillation is that the probability of success is not equal to 1 in this case, assuming a limited supply of free entangled states.\\
While in original entanglement distillation two copies of the same state were subjected to local measurements at a time, here it is one copy of the bound entangled state and one copy of the free entagled one. Hence, if the measurement outcome is unsatisfactory, the used free entangled state is lost.\\
Nevertheless, this concept allows bound entanglement to be much more useful than it might seem at first glance, even if it still does not compare to free entanglement.\\ \\
It is now obvious, how knowledge about different kinds of entangled states can be useful in order to be able to understand quantum information theory and to practically apply it. Therefore it is necessary to find various means to detect entanglement and to find out which of these are useful in respect to the different types of entanglement.\\ 
\newpage

\section{Detecting Entanglement}
From now on, entanglement shall be viewed from a more theoretical and mathematical point of view.\\
In order to detect entanglement in quantum states, one obviously first needs to define it properly.\\

\subsection{Defining Entanglement}
A bipartite pure state $\left|\Psi\right\rangle$ is called separable iff it can be written as a single tensor product of states in the subsystems A and B (that is, if its Schmidt rank equals one):
\begin{equation}\label{eq:sepvec} \left|\right.\Psi\left.\right\rangle = \left|\right.\Psi^{A}\left.\right\rangle \otimes \left|\right.\Psi^{B}\left.\right\rangle \end{equation}
Every nonseparable state vector is called entangled and has the form
\begin{equation} \left|\right.\Psi\left.\right\rangle = \sum_{i=1}^{d_{A}} \sum_{j=1}^{d_{B}} c_{ij} \left|\right.\Psi_{i}^{A}\left.\right\rangle \otimes \left|\right.\Psi_{j}^{B}\left.\right\rangle \end{equation}
with at least two nonzero complex coefficients $c_{ij}$.\\
Equivalently, a pure state is called separable iff its reduced density matrices correspond to pure states.\\
A state is maximally entangled, if it is pure and its reduced density matrices are maximally mixed. Equivalently, a pure state is maximally entangled iff it has full Schmidt rank and each term is equally weighted.\\ \\
While the situation is rather simple for pure states, analysing mixed states can be very difficult due to the nonuniqueness of decompositions of density matrices into pure states. A mixed state is called separable iff it can be written in the form
\begin{equation}\label{eq:sepdm} \rho = \sum_{i=1}^{d_{A}} \sum_{j=1}^{d_{B}} p_{ij} \ \rho_{i}^{A} \otimes \rho_{j}^{B} \end{equation}
where $d_{A}$ and $d_{B}$ are the dimesions of the subspaces, the $\rho_{i}^{A}$ and $\rho_{j}^{B}$ are density matrices of the respective subspaces and the $p_{ij}$ are probabilities, such that
\begin{equation} p_{ij} \geq 0, \ \ \ \mathrm{and} \ \ \ \sum_{i,j}p_{ij} = 1\end{equation}
For reasons of convenience, the indices can be chosen such that eq. (\ref{eq:sepdm}) assumes the form
\begin{equation} \rho = \sum_{i} p_{i} \ \rho_{i}^{A} \otimes \rho_{i}^{B} \end{equation}
For mixed states, the form of an entangled state cannot be explicitly formulated other than by saying that it cannot be written in the form (\ref{eq:sepdm}). That is why it is rather difficult to decide, if a given state is separable or entangled.\\

\subsection{Pure States}
As mentioned, for pure states, detecting entanglement is a rather simple task and can be done in various ways.
\begin{enumerate}
	\item According to the definition of entangled pure states (\ref{eq:sepvec}), a separable state's Schmidt rank equals one. Hence, iff any state has a Schmidt rank greater than one, it is entangled.
	\item From the definition of entanglement follows, that the composite system contains more information (i.e. purity) than the subsystems. In particular, the reduced density matrices of a pure state are pure states themselves iff the state is separable. Consequently, a pure state $\rho$ is separable iff one of the following equivalent statments is true:
	\[ \mathrm{Tr}((\rho^{A})^{2}) = 1 \]
	\begin{equation}\label{pure_entr} S(\rho^{A}) > 0 \end{equation}
	where
	\begin{equation}\label{vn_entropy} S(\rho) = - \mathrm{Tr}(\rho \log \rho)\end{equation}
	is the von Neumann entropy, which is a measure for the mixedness of a quantum state.
	\item Of course every separability criterion for general states can be applied to a pure state as well, although this is generally not the most economic way to study pure states. Still, iff a pure state is separable, it satisfies any of the separability criteria that will be discussed in this section.\\
\end{enumerate}

\subsection{Positive and Completely Positive Maps\label{pms}}
A map $\Lambda: \mathcal{B} \rightarrow \mathcal{B}$ is called a positive map (PM) iff it maps all positive operators $\rho$ onto positive operators
\begin{equation} \rho \geq 0 \Rightarrow \Lambda(\rho) \geq 0 \ \ \ \forall \rho \in \mathcal{B} \end{equation}
A PM is called completely positive (CP) iff it remains positive under extensions to all higher dimensions
\begin{equation} \rho \geq 0 \Rightarrow (\Lambda\otimes\mathds{1}_{d}) (\rho) \geq 0 \ \ \ \forall \rho \in \mathcal{B}\otimes\mathds{C}^{d}, \ \ \ \forall d \in \mathds{N} \end{equation}
Interestingly, not all PMs have this property, which allows maps that are positive but not CP to detect entangled states.\\
It is easy to see that any PM leaves a separable state $\rho$ positive, since
\begin{equation} (\Lambda\otimes\mathds{1}) (\rho) = (\Lambda\otimes\mathds{1}) \left(\sum_{i} p_{i} \ \rho_{i}^{A} \otimes \rho_{i}^{B}\right) = \sum_{i} p_{i} \ \Lambda (\rho_{i}^{A}) \otimes \rho_{i}^{B} \geq 0 \end{equation}
which is not true for general states. In fact, for each entangled state $\rho$ there exists a PM $\Lambda$ such that\cite{horo_pms}
\begin{equation} (\Lambda\otimes\mathds{1}) (\rho) \ngeq 0 \end{equation}
Conversely, a state $\rho$ is separable iff it satisfies
\begin{equation} (\Lambda\otimes\mathds{1}) (\rho) \geq 0 \end{equation}
for all PMs $\Lambda$.\\ \\
Although this is a necessary and sufficient separability criterion, its strength cannot be used to its full extent, since there is no way to apply it to any given state. Presently, knowledge about non-CP PMs is rather limited. However, as far as known, non-CP PMs can be used to formulate necessary separability criteria.\\

\subsubsection{PPT (Peres-Horodecki) Criterion\label{ppt}}
The probably ''strongest`` PM (in the sense of being able to detect most entangled states) is the transposition $T$. Due to the arguments given in the previous subsection, the partial transposition $T\otimes\mathds{1}$ of a seperable state is positive, while there are entangled states that behave differently (as mentioned in section \ref{be}). Hence, a state with nonpositive partial transposition (NPT) has to be entangled, while a state with positive partial transposition (PPT) can be either separable or (bound) entangled\cite{peres_ppt}. This is called the PPT-criterion (or Peres-Horodecki criterion) of separability.\\
In low dimensions, this criterion is much stronger. For systems of two qubits\cite{2x2pm} and systems of one qubit and one qutrit\cite{2x3pm} the transposition is the only relevant PM, since here all PMs $\Lambda$ are decomposable, i.e. can be written in the form
\begin{equation}\label{mapdecomp} \Lambda = \Lambda_{1}^{CP} + \Lambda_{2}^{CP} T \end{equation}
where the $\Lambda_{i}^{CP}$ are CP maps.\\ \\
From this follows that any state that is nonpositive under any of these maps must also be NPT, since
\begin{equation} (\Lambda\otimes\mathds{1}) (\rho) = (\Lambda_{1}^{CP}\otimes\mathds{1}) (\rho) + (\Lambda_{2}^{CP}\otimes\mathds{1}) (\rho^{T_{A}}) \end{equation}
can only be nonpositive if $\rho^{T_{A}}$ is.\\
Consequently, in these cases the PPT criterion is both necessary and sufficient for separability, thus providing a very simple procedure to determine without a doubt if any given state in these dimensions is separable or entangled.\\

\begin{center}\emph{Partial Transposition of Expansions of Given States}\end{center}
Consider states $\rho'$ on the Hilbert space $\mathcal{H} = \mathcal{H}_{A}\otimes\mathcal{H}_{B}\otimes\mathcal{H}_{aux} = \mathcal{H}_{A}^{\otimes k}\otimes\mathcal{H}_{B}^{\otimes l}$, such that $\rho$´ is an extension of $\rho$ (i.e. $\mathrm{Tr}_{\mathcal{H}_{aux}}(\rho') = \rho$) that is symmetrical under exchange of any copies of the spaces $\mathcal{H}_{A}$ and $\mathcal{H}_{B}$.\\
If $\rho$ is seperable, such an expansion is of the form
\begin{equation} \rho' = \sum_{i} p_{i} \ (\left|\Psi^{A}_{i}\right\rangle\left\langle\Psi^{A}_{i}\right|)^{\otimes k}\otimes(\left|\Psi^{B}_{i}\right\rangle\left\langle\Psi^{B}_{i}\right|)^{\otimes l} \end{equation}
and hence is positive under any partial transposition.\\
Consequently, if there exists an expansion $\rho'$ of a state $\rho$ that satisfies the above symmetry conditions and has nonpositive partial transposition, $\rho$ is entangled\cite{expppt}.\\ \\
Note that this is a stronger criterion than the normal PPT-criterion, since if a state is NPT, there automatically exists a NPT expansion, while the converse of this statement is not true (as can be seen from examples in ref. \cite{expppt}).\\
In order to determine if a given state is entangled, this criterion can by used stepwise. If the partial transposition itself is positive, it can be expanded to a higher dimensional Hilbert space. If this expansion is still positive, it can be expanded further, and so on. Means to find suitable expansions have been shown in ref. \cite{expppt}.\\

\subsubsection{Reduction Criterion}
Another example for non-CP PMs is the reduction criterion\cite{reduction}, which consists of applying the positive map
\begin{equation} \Lambda(\sigma) = \mathds{1} \mathrm{Tr}(\sigma) - \sigma \end{equation}
to one of the subsystems, resulting in the separability criteria
\begin{equation}\begin{split} (\Lambda\otimes\mathds{1}) (\rho) = \mathds{1}\otimes\rho_{B} - \rho \geq 0 \\ (\mathds{1}\otimes\Lambda) (\rho) = \rho_{A}\otimes\mathds{1} - \rho \geq 0 \end{split}\end{equation}
Although the reduction criterion is necessary and sufficient for separability for systems in $2\otimes2$ and $2\otimes3$, it follows from the discussion in (\ref{ppt}) that it is a weaker criterion than the PPT criterion, since it does not detect all NPT states in higher dimensions. However, violation of the reduction criterion is equivalent to distillability via a special class of distillation protocols, that is a straightforward generalisation of the usual $2\otimes2$ dimensional one. In particular, any state violating it can be distilled (while a state satisfying it may still be distillable by a different protocol) and is hence free entangled.\\

\subsection{Cross-Norm- and Realignment Criterion\label{ccn-crit}}
The cross-norm criterion of separability\cite{crossnorm1} states, that a state $\rho$ is separable iff $\left\|\rho\right\|_{\gamma} = 1$, where the cross-norm $\left\|.\right\|_{\gamma}$ of a density matrix is defined by
\begin{equation} \left\|t\right\|_{\gamma} = \inf_{\{u_{i}, v_{i}\}} \sum_{i} \left\|u_{i}\right\|_{1}\left\|v_{i}\right\|_{1} \end{equation}
where the infimum is taken over all sets of weighted density matrices $\{u_{i}, v_{i}\}$ satisfying $t = \sum_{i} u_{i} \otimes v_{i}$.\\
This norm is also often refered to als the greatest cross norm, since it majorises any subcross norm (a norm is said to be subcross, if $\left\|a \otimes b\right\| \leq \left\|a\right\| \left\|b\right\|$ and is called cross if the equality holds).\\
Although the cross-norm criterion is necessary and sufficient for separability, it is in general not applicable, since the infimum is hard to compute. However, it implies a weaker criterion, that is not sufficient for separability, but easily computable.\\ \newpage

\begin{center}\emph{Realignment criterion}\end{center}
It follows from the cross-norm criterion\cite{ccn}, that if a state $\rho$ is separable, then
\begin{equation} \left\|\rho_{R}\right\|_{1} \leq 1\end{equation}
where the realigned density matrix $\rho_{R}$ is defined by
\begin{equation} \rho_{R} = \sum_{i,j,k,l} \left\langle i,k\right| \rho \left|j,l\right\rangle \ \left|i,j\right\rangle\left\langle k,l\right| \end{equation}
Note that despite the formal similarity between the definitions of the realigned density matrix and the partially transposed density matrix, they are completely different objects. In particular, $\rho_{R}$ is not hermitian. This realignment criterion is -- since it is a consequence of the cross-norm criterion -- also often referred to as the computable cross-norm (CCN) criterion.\\
The realignment criterion was proven\cite{further_ccn} to be independent from (i.e. neither stronger nor weaker than) the PPT criterion, as it is capable of detecting certain PPT entangled states, whereas it is not sufficient in the $2\otimes2$ case\cite{further_ccn} (there are however whole sets of states, for which it is sufficient, such as all pure states or states of $2\otimes2$ systems with maximally mixed subsystems).\\
Due to the mathematical similarity between the realigned density matrix and the partially transposed density matrix, it seems reasonable that an expansion to higher spaces could lead to a more powerful hierachical set of separability criteria, like it does for the PPT criterion.\\ 

\subsection{Majorisation Criterion}
A completely different approach can be given by considering the mixedness of a given quantum state and its reductions (i.e. reduced density matrices), thus generalising the entropic separability criterion for pure states (\ref{pure_entr}). Many criteria were formulated using the von Neumann entropy of both these objects and relating them to each other, it turns out however, that a stronger separability criterion can be given.\\
If a state $\rho$ is separable, then is satisfies\cite{major}
\begin{equation}\begin{split} \lambda(\rho) \prec \lambda(\rho_{A}) \\ \lambda(\rho) \prec \lambda(\rho_{B})\end{split}\end{equation}
where $\rho_{A,B}$ are the reduced density matrices, $\lambda(\sigma)$ is the vector of $\sigma'$s eigenvalues (where there are zeroes appended to the $\lambda(\rho_{A,B})$, such that all vectors have the same dimension). A vector $x = (x_{1}, ..., x_{d})$ is said to be majorised by a vector $y = (y_{1}, ..., y_{d})$ (i.e. $x \prec y$) if
\begin{equation} \sum_{i=1}^{k}x^{\downarrow}_{i} \leq \sum_{i=1}^{k}y^{\downarrow}_{i} \ \ \ \forall k\end{equation}
where $x^{\downarrow}$ is the vector $x$ with sorted components, i.e. $x_{1} \leq ... \leq x_{d}$ (and analogously for $y$).\\ \\
Note that if a state $\rho$ satisfies the majorisation criterion, it follows that it also satisfies weaker entropic criteria, for example
\begin{equation}\begin{split} S(\rho) \geq S(\rho_{A}) \\ S(\rho) \geq S(\rho_{B}) \end{split}\end{equation}
where $S(\sigma)$ is the von Neumann entropy (\ref{vn_entropy}).\\

\subsection{Range Criterion}
A separability criterion seemingly independant from all the ones mentioned above can be given via the range of density matrices and their partial transpo\-sitions\cite{range}. The range of $\rho$ is defined as the set of all vectors $\left|\Psi\right\rangle$ for which there is another vector $\left|\Phi\right\rangle$ such that
\begin{equation} \left|\Psi\right\rangle = \rho \left|\Phi\right\rangle \end{equation}
If a state $\rho$ on a Hilbert space $\mathcal{H}$ with $\mathrm{dim}\mathcal{H} = m$ is separable, then there exists a set of product vectors $\left\{\psi_{i}\otimes\phi_{j}\right\}, \left\{i,j\right\} \in I$ (where $I$ is a set of N pairs of indices with $N = \#I \leq m^{2}$) and probabilities $\{p_{i,j}\}$ such that
\begin{enumerate}
	\item the ensembles $\{\psi_{i}\otimes\phi_{j}, p_{i,j}\}$, $\{\psi_{i}^{*}\otimes\phi_{j}, p_{i,j}\}$ and $\{\psi_{i}\otimes\phi_{j}^{*}, p_{i,j}\}$ correspond to the matrices $\rho$, $\rho^{T_{A}}$ and $\rho^{T_{B}}$
	\item the ranges of $\rho$, $\rho^{T_{A}}$ and $\rho^{T_{B}}$ are spanned by the vectors $\{\psi_{i}\otimes\phi_{j}\}$, $\{\psi_{i}^{*}\otimes\phi_{j}\}$ and $\{\psi_{i}\otimes\phi_{j}^{*}\}$, respectively
\end{enumerate}
where the asterisk denotes complex conjugation.\\
It was shown\cite{range} that this criterion is independant from the PPT criterion, as there are PPT entangled states that violate the range criterion and there also are NPT states satisfying it.\\

\subsection{Entanglement Witnesses}
\begin{figure}[ht!]\centering\includegraphics[width=80mm]{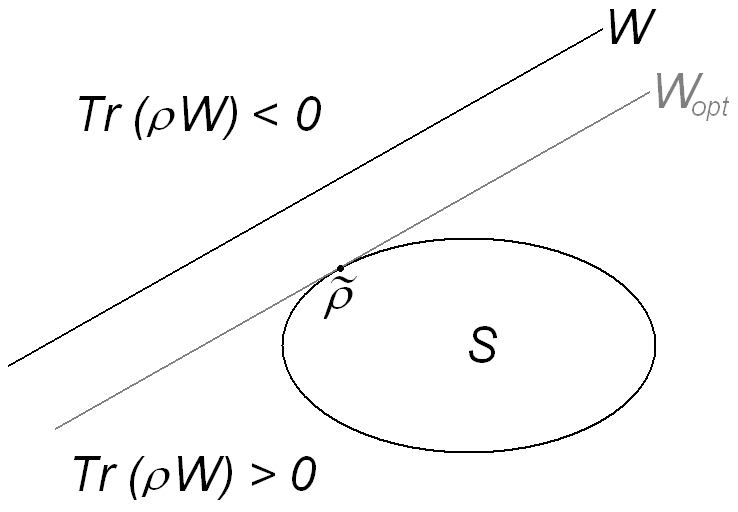}\caption{Illustration of entanglement witnesses}\label{fig:ew}\end{figure}

From the definition of seperable states
\begin{equation} \rho = \sum_{i} p_{i} \left|\Psi_{i}^{A}\right\rangle\left\langle\Psi_{i}^{A}\right|\otimes\left|\Psi_{i}^{B}\right\rangle\left\langle\Psi_{i}^{B}\right| \end{equation}
it follows immediately that the set of separable states $S$ is a convex and closed subset of the set of all states on $\mathcal{H} = \mathcal{H}_{A}\otimes\mathcal{H}_{B}$. Also, of course a set containing only one state is convex, compact and closed.\\
If there are two convex and compact sets of which at least one is closed, then there exists a hyperplane that separates these sets from each other, i.e. one set lies completely on one side of the plane, while the other set lies completely on the other side. This is a corollary from the Hahn-Banach theorem\cite{hahnbanach}.\\
From this now follows the entanglement witness theorem, which states that for each entangled state $\rho$ there exists a hermitian operator $W$ such that
\begin{equation}\label{ew0} \mathrm{Tr}(\rho W) < 0 \end{equation}
while
\begin{equation}\label{ew1} \mathrm{Tr}(\sigma W) \geq 0 \ \ \ \forall \ \sigma \in S \end{equation}
This operator is called an entanglement witness. If in addition there exists a separable state $\tilde{\rho}$ such that
\begin{equation}\label{ew2} \mathrm{Tr}(\tilde{\rho} W) = 0 \end{equation}
then $W$ is called an optimal entanglement witness\cite{horo_pms}. \\ \\
This is a direct consequence of the above statement, since all states $\rho$ with $\mathrm{Tr}(\rho W) = 0$ form a hyperplane (remember that $\mathrm{Tr}(A B) = \left\langle A | B \right\rangle$ is the scalar product in the Hilbert-Schmidt space for hermitian operators $A$ and $B$), and all states $\rho_{+}$ satisfying $\mathrm{Tr}(\rho_{+} W) > 0$ are located on one side of it, while all states $\rho_{-}$ with $\mathrm{Tr}(\rho_{-} W) < 0$ are located on the other side. If $W$ is an optimal entanglement witness (often denoted by $W_{opt}$), the corresponding hyperplane is a tangent plane to S (for illustration see Fig. \ref{fig:ew}). Note that $W$ and $\alpha W$ are the same witness for any complex number $\alpha$.\\
Of course, optimal entanglement witnesses are of high interest, since they can detect more entangled states than nonoptimal ones and give a deep insight into the geometrical structure of a Hilbert space. In particular, the set of separable states $S$ is fully characterised by the set of all optimal entanglement witnesses (since these form the border of $S$\cite{bertl_narn_thirring}).\\
Obviously, it is rather difficult to check wether an arbitrary operator satisfies condition (\ref{ew1}). However, in some cases there are means to check this. \\

\begin{center}\emph{Is a given operator an entanglement witness?}\end{center}
In order to see if a given operator is indeed an entanglement witness, and in particular satisfies condition (\ref{ew1}), it is useful to investigate decompositions of general separable states into generlaised Gell-Mann matrices $\lambda_{i}$ (or any other basis of the respective Hilbert space, see \ref{decomp}). Any state can be decomposed in the following way:
\begin{equation} \sigma = \frac{1}{d^2}\left(\mathds{1}\otimes\mathds{1} + \sum_{i=1}^{d^{2}-1} a_{i}\lambda_{i}\otimes\mathds{1} + \sum_{i=j}^{d^{2}-1}b_{j}\mathds{1}\otimes\lambda_{j} + \sum_{i,j=1}^{d^{2}-1}c_{i,j}\lambda_{i}\otimes\lambda_{j}\right) \end{equation}
with adequate bounds on the coefficients $a_{i}$ and $b_{i}$ (depending on the dimension $d$). The bounds on $c_{i,j}$ for separable states are much tighter than those for general states. Thus, the quantity $\mathrm{Tr}(\sigma W)$ can be computed for all separable states, allowing to check if it is positive.\\

\begin{figure}[ht!]\centering\includegraphics[width=80mm]{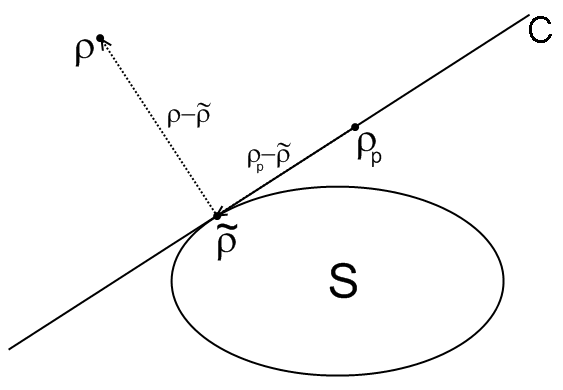}\caption{Illustration of an optimal geometric entanglement witness}\label{fig:gew}\end{figure}

\begin{figure}[ht!]\centering\includegraphics[width=80mm]{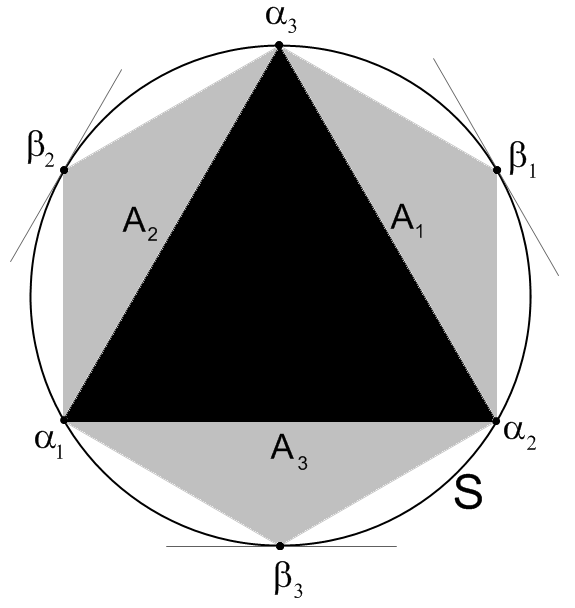}\caption{Illustration of the procedure to construct the set of separable states (inside-out-shifting)}\label{fig:gew_sep}\end{figure}

\begin{center}\emph{Geometric entanglement witnesses}\end{center}
A geometrically very intuitive way of constructing entanglement witnesses is the method of geometric entanglement witnesses\cite{gew}. Given an entangled state $\rho$ the operator
\begin{equation}\label{gew} C = \tilde{\rho} - \rho - \left\langle\tilde{\rho}|\tilde{\rho} - \rho\right\rangle\mathds{1} \end{equation}
is an entanglement witness detecting $\rho$ for certain states $\tilde{\rho}$. In particular, $C$ is an optimal entanglement witness iff $\tilde{\rho}$ is the separable state closest to $\rho$ in the Hilbert-Schmidt metric (methods to find the closest separable states are discussed in section \ref{hilbertschmidt}). This can be easily understood regarding the illustration in Fig. \ref{fig:gew}: Due to the construction of $C$, the corresponding hyperplane is orthogonal to the line between $\rho$ and $\tilde{\rho}$ and also contains $\tilde{\rho}$ itself, since
\begin{equation}\begin{split} 0 = \mathrm{Tr}(\rho_{p} C) &= \mathrm{Tr}(\rho_{p} (\tilde{\rho} - \rho - \left\langle\tilde{\rho}|\tilde{\rho} - \rho\right\rangle\mathds{1})) \\
&= \left\langle\rho_{p}|\tilde{\rho}-\rho\right\rangle - \left\langle\tilde{\rho}|\tilde{\rho}-\rho\right\rangle \\
&= \left\langle\rho_{p}-\tilde{\rho}|\tilde{\rho}-\rho\right\rangle \end{split}\end{equation}
where the last equality holds since $\rho_{p}$ per definition lies on the plane (and in particular if $\rho_{p} = \tilde{\rho}$). \\ \\
Geometric entanglement witnesses can be used very effectively to iteratively construct the complete set of separable states\cite{gewshift} (as illustrated in Fig. \ref{fig:gew_sep}). To start, one needs a number of separable states $\alpha_{1, ..., n}$, whose convex hull evidently forms a polytope containing only separable states (black triangle in Fig. \ref{fig:gew_sep}). Each of this polytope's borderplanes can be described by a geometric operator $A_{i}$ (i.e. a operator of the form \ref{gew}) that is not an entanglement witness, since its expectation value is not positive for all separable states. These operators can now be shifted towards the edge of the set of separable states $S$, whilst observing the entanglement witness criterion eq. (\ref{ew1}) (by means of decomposition coefficients, as discussed above) until they become optimal entanglement witnesses corresponding to tangent hyperplanes of $S$ (grey lines in Fig. \ref{fig:gew_sep}) which contain a new separable state $\beta_{i}$ each. The states $\alpha_{i}$ and $\beta_{i}$ together now form a new polytope, containing more separable states.\\
This procedure is called inside-out-shifting and can be repeated arbitrarily often. If $S$ is a polytope, then the procedure is finite and ends as soon as all resulting entanglement witness planes contain more than one separable state, i.e. form the borderplanes of $S$. If $S$ has a more complex shape (as in Fig. \ref{fig:gew_sep}), the procedure can be repeated over and over until the desired precision is achieved.\\
In contrast to inside-out-shifting, the same result can be obtained by outside-in-shifting. In this case, one starts with a set of nonoptimal geometric entanglement witnesses and shifts them towards the set of separable states until they become optimal.\\

\begin{center}\emph{Relations between entanglement witnesses and other separability criteria}\end{center}
Since entanglement witnesses are very general objects, many other separability criteria can be written as such.\\
A very intuitive relation is the one between entanglement witnesses and positive maps. Since these both are necessary and sufficient separability criteria, there has to be both an (optimal) entanglement witness and a positive map for each nonseparable state to be detected, between which one might intuitively conjecture that there exists a kind of isomorphism (of course, for each PM there need to be several entanglement witnesses due to the geometry of the set of states detected by the PM, which in general cannot be described by a single hyperplane). This isomorphism is called the Jamiolkowski-isomorphism\cite{jamiolkowski1, jamiolkowski2}. For each non-CP PM $\Lambda$ there exists a family of entanglement witnesses
\begin{equation} W = \left(\Lambda\otimes\mathds{1}\right) (\left|\Psi\right\rangle\left\langle \Psi\right|) \end{equation}
where $\left|\Psi\right\rangle$ is any maximally entangled state.\\
Evidently, this makes any separability criterion formulated via positive maps (in particular the PPT criterion and the reduction criterion) possible to be considered as a set of entanglement witnesses.\\

\begin{center}\emph{Structure of entanglement witnesses}\end{center}
Since entanglement witnesses are a very general class of operators, their mathematical structure is rather general as well. Obviously, a product operator cannot be an entanglement witness, but already a sum of two product operators can be\cite{bertl_narn_thirring}.\\
In order to yield positive and negative expectation values for different states, an entanglement witness evidently needs to be an indefinite operator. However, from the positivity for all separable states it follows, that all diagonal elements in the computational basis have to be positive semidefinite.\\

\subsection{Entanglement Measures}
In the following section, measures for entanglement will be discussed. Of course, if a quantum state possesses any nonzero amount of entanglement (in any suitable measure), it needs to be entangled. Hence, entanglement measures can, in a way, also be used to detect entanglement, although these are usually much more difficult to compute than the separability criteria discussed above. Also, most of the following entanglement measures are quantisations of existing separability criteria (as will be seen), so that the procedure of detecting entanglement by means of a measure is fully equivalent to using the corresponding separability criterion in the first place.\\ \newpage

\section{Entanglement Measures}
Trying to quantify entanglement, one is immediately confronted with an important question: What properties does an entanglement measure need to have in order to be useful? This question cannot be answered completely objectively. While some properties seem to be necessary, others are only needed in order to preserve intuitive pictures which are not necessarily correct.\\
Usually, the following catalog of properties is wished to be fulfilled by a ``good'' entanglement measure $E(\rho)$\cite{bruss}:
\begin{enumerate}
	\item $\rho$ is separable $\Leftrightarrow E(\rho) = 0$\\
	This is a very intuitive requirement. A measure should only yield no entanglement at all for separable states and should not yield any entanglement for any separable state.
	\item $\rho$ is maximally entangled $\Leftrightarrow E(\rho) = \max E(\omega)$
	where the maximum is taken over all states $\omega$. An entanglement measure should yield its maximal value for maximally entangled states and for maximally entangled states only.
	\item No increase under LOCC: $E(\rho) \geq E(\Lambda_{LOCC}(\rho))$\\
	Since entangled states cannot be created by local operations and classical communications (LOCC), it seems reasonable, that the amount of entanglement in a quantum state cannot be increased by LOCC either.
	\item Invariance under local unitary operations: $E(\rho) = E(U_{1}\otimes U_{2} \rho U_{1}^{\dagger}\otimes U_{2}^{\dagger})$\\
	Since unitary operations can be regarded as changes of the basis, this is equivalent to the statement that the entanglement should not depend on the choice of basis. If this requirement is fulfilled, it induces equivalence classes of states containing the same amount of entanglement and being linked to each other via unitary transformations. This condition is in short also referred to as unitary invariance, omitting the ``local''
	\item Continuity: $||\rho_{1}-\rho_{2}|| \rightarrow 0 \Rightarrow E(\rho_{1}) - E(\rho_{2}) \rightarrow 0$\\
	Any entanglement measure should be continuos on the Hilbert space of states, so that the entanglement cannot change dramatically for infinitesimal changes of the state.
	\item Convexity: $E(\lambda \rho_{1} + (1-\lambda) \rho_{2}) \leq \lambda E(\rho_{1}) + (1-\lambda) E(\rho_{2})$\\
	for $0 < \lambda < 1$. The entanglement of any convex combination of two states should not exceed the convex sum of their single entanglements. This requirement is very intuitive in a geometric point of view, since a state's distance from the set of separable states $S$ intuitively corresponds to an entanglement measure itself. Now, since $S$ is convex, any convex combination of two states $\notin S$ can only be closer to $S$ than its components. Also, it seems obvious that the more mixed a state is, the less entanglement it can contain, since otherwise entanglement could be created simply by mixing states.
	\item Additivity: $E(\rho_{1}\otimes\rho_{2}) = E(\rho_{1})+E(\rho_{2})$\\
	Any two (or more) states should always contain exactly the same entanglement that is contained in each of them taken together, since these states can easily be separated from each other and be used to perform quantum tasks exploiting entanglement, the possibility of which should be contained in an entanglement measure.\\
	Often only a weaker form of this property is required, namely additivity for equal states and subadditivity for inequal states:\\
	$E(\rho^{\otimes n}) = n E(\rho)$ and	$E(\rho_{1}\otimes\rho_{2}) \leq E(\rho_{1}) + E(\rho_{2})$
	\item Normalisation: $0 \leq E(\rho) \leq \log d$ where d is the dimension of the Hilbert space. Usually, when dealing with only one particular Hilbert space, the logarithmic base is chosen to be $d$, such that $0 \leq E(\rho) \leq 1$.	In general, the choice of the logarithmic base depends on the units in which the entanglement is measured, so called dits of entanglement (edits). In most cases, the base is chosen to be two, such that the entanglement is measured in ebits.
	\item For an entanglement measure to be useful, there needs to be an operational way of computation for all states. This turns out to be the main problem very often.
\end{enumerate}
It turns out, that finding an entanglement measure that fulfills all of the above criteria is extremely difficult, which is why there are many candidates, each of which does not satisfy all of the criteria.\\

\subsection{Pure States\label{puremeasure}}
For pure states, it is very easy to quantify the separability criteria discussed in the previous section. Since entanglement can be seen as the amount of information that is contained in a system, but not in its subsystems taken together (i.e. the information that is lost when viewing not the composite system but its subsystems), the mixedness of the subsystems seems an appropriate measure for entanglement:
\begin{equation} E(\rho) = S(\rho^{A}) = S(\rho^{B})\end{equation}
where $\rho^{A,B}$ are the reduced density matrices of $\rho$ and $S$ is any suitable entropy function, for example the von-Neumann-entropy $S(\sigma) = - \mathrm{Tr}(\sigma \log \sigma)$ or the linear entropy $S(\sigma) = \frac{d}{d-1}(1 - \mathrm{Tr}(\sigma^{2}))$.\\
This family of measures satisfies all of the properties discussed above and hence seems to be a very good measure for entanglement. However, it can only be applied to pure states.\\

\subsection{Bell Inequalities\label{bell}}
Historically, the first discrimination of quantum correlations as opposing to classical correlations was found by John S. Bell\cite{bell}, who formulated a class of correlation-inequalities -- Bell inequalities (BI) -- that any local realistic theory necessarily has to satisfy. The terms ``local'' and ``realistic'' are here used in the sense of Einstein, Podolsky and Rosen\cite{epr}:\\
\emph{Locality:} Any physical theory should be compatible with the theory of special relativity and especially not allow any faster than light transmission of information.\\
\emph{Realism:} Any physical quantity which can be predicted with certainty and without disturbing the system, corresponds to an element of reality. Hence, if a quantity is considered to be physically real, its value should exist independently of any measurement.\\\\
Obviously, quantum mechanics violates these criteria (to be more precise: depending on the choice of interpretation, it violates at least one of them. According to the widely accepted Kopenhagen interpretation, both are violated, while for example the Bohmian interpretation\cite{bohm} is only nonlocal but realistic) and also violates Bell's inequalities.\\
In accordance with the commonly used terminology, a quantum state is called nonlocal if it violates any Bell inequality, while else (if it satisfies all Bell-inequalities) it is called a local state. This however does not necessarily require it to be ``nonlocal'' in the actual sense, since the violation can also be explained by a local but nonrealistic model.\\
Note that there obviously have to be entangled states, which do not violate any Bell inequality, since these represent an upper bound to correlations and do not distinguish between classical correlations (which result from mixing of pure product states) and quantum correlations. If a state possesses a certain amount of quantum correlations (i.e. entanglement) and just a very little amount of classical correlations, it may not violate any BI, since its total correlations are well within range of local realism.\\

\begin{center}\emph{Local Hidden-Variable Models}\end{center}
Any local realistic theory can be formulated via local hidden variables (LHV), i.e. a set of parameters $\lambda$ that completely determine the outcome of any given measurement and are local (in the sense of there being no interaction or dependance between the parameters for systems that are spacially separated from each other), while the parameters themselves do not have to be accessable by any experiment.\\
It is obvious, that any quantum product state can be described by LHVs, but the question remains if the same holds for entangled states. As mentioned above, there are entangled states that violate Bell's inequalities (nonlocal states) and such that do not (local states). For pure states the situation is rather simple, as any entangled pure state is nonlocal and incompatible with all LHV models\cite{gisin_pure, popescu_rohrlich}. For mixed states, the distinction is not that easy. Clearly, nonlocal states can never be fully reproduced by LHV models, however, surprisingly, there are local states for which there does not exist a suitable LHV-description\cite{mermin_lhv}, while there also are entangled states that can be modeled by LHVs\cite{werner_lhv}.\\

\begin{center}\emph{The CHSH-Inequality}\end{center}
The by far most commonly used Bell inequality is the one for systems of $2\otimes2$ dimensions derived by Clauser, Horne, Shimony and Holt\cite{chsh} (CHSH inequality). In a system of 2 qubits, the expectation value of a bilateral (e.g. spin-) measurement along the directions $\vec{a}$ in the first system $\vec{b}$ in the second can be defined in a LHV model by
\begin{equation} E(\vec{a},\vec{b}) = \int A(\vec{a},\lambda) B(\vec{b}, \lambda) \rho(\lambda) d\lambda \end{equation}
where $\rho(\lambda)$ is any distribution function for the hidden value $\lambda$ and $A(\vec{a},\lambda)$ and $B(\vec{b},\lambda)$ are the functions determining the (normalised) measurement results in both systems, depending on the direction of measurement and the hidden parameter.\\
Using this definition, the CHSH inequality can be found, starting with the identity
\begin{equation}
	E(\vec{a_{1}},\vec{b_{1}}) - E(\vec{a_{1}},\vec{b_{2}}) = E(\vec{a_{1}},\vec{b_{1}}) (1 \pm E(\vec{a_{2}},\vec{b_{2}})) - E(\vec{a_{1}},\vec{b_{2}}) (1 \pm E(\vec{a_{2}},\vec{b_{1}}))
\end{equation}
it follows by using the triangle inequality
\begin{equation}\begin{split}
	\left|E(\vec{a_{1}},\vec{b_{1}}) - E(\vec{a_{1}},\vec{b_{2}})\right| & \leq \left|E(\vec{a_{1}},\vec{b_{1}})(1 \pm E(\vec{a_{2}},\vec{b_{2}}))\right| + \left|E(\vec{a_{1}},\vec{b_{2}})(1 \pm E(\vec{a_{2}},\vec{b_{1}}))\right| \\
	& \leq \left|1 \pm E(\vec{a_{2}},\vec{b_{2}})\right| + \left|1 \pm E(\vec{a_{2}},\vec{b_{1}})\right|
\end{split}\end{equation}
where the second line follows from the normalisation $-1 < E(\vec{a_{i}},\vec{b_{j}}) < 1$. The inequality is strongest when the sign is chosen such that
\begin{equation}\label{chsh} I = \left|E(\vec{a_{1}},\vec{b_{1}}) - E(\vec{a_{1}},\vec{b_{2}})\right| + \left|E(\vec{a_{2}},\vec{b_{2}}) + E(\vec{a_{2}},\vec{b_{1}})\right| \leq 2\end{equation}
This is the CHSH inequality.\\
Evidently, for local realistic theories, the maximal value (maximised over all suitable states and all directions $\vec{a_{i}}$ and $\vec{b_{i}}$) for $I$ is 2, while the maximal possible value for any theory is 4 (since the single expectation values cannot exceed $\pm 1$). Quantum mechanics however predicts a maximal value of $2 \sqrt{2}$, which is reached by maximally entangled states.\\\\
Entangled pure two qubit states always violate this inequality. In an appropriate basis they can always be written in the form $\left|\Psi\right\rangle = \alpha \left|\uparrow\uparrow\right\rangle + \beta \left|\downarrow\downarrow\right\rangle$ (with $\alpha, \beta \in \mathds{R}$ and $\alpha^2 + \beta^2 = 1$), then the maximal value (optimised over all angles) of the Bell parameter is $I = 2\sqrt{1+4\alpha^2\beta^2}$\cite{popescu_rohrlich}.\\
Since mixed states are more complicated to deal with, it is fortunate that a constructive way was found\cite{horo_chsh} to compute the maximal value of the Bell parameter for any two qubit state. Any two qubit state $\rho$ can be written as
\begin{equation} \rho = \frac{1}{4}(\mathds{1} + \vec{a} \vec{\sigma}\otimes\mathds{1} + \mathds{1}\otimes\vec{b}\vec{\sigma} + \sum_{m,n=1}^{3} t_{mn} \sigma_{m}\otimes\sigma_{n}) \label{bloch2x2}\end{equation}
where the $t_{mn}$ form a real three by three matrix $T_{\rho}$. The maximal value for the Bell parameter then can be written as 
\begin{equation} I_{max} = 2(u_{1} + u_{2}) \label{2x2bell}\end{equation}
where $u_{1,2}$ are the two larger eigenvalues of the matrix $T^{T}_{\rho}T_{\rho}$.\\\\
The CHSH inequality is -- in $2\otimes2$ dimensions -- optimal in the sense that there is no other Bell inequality that is either more resistant to noise or yields a higher maximal violation\cite{gen_chsh}.\\

\begin{center}\emph{Bell Inequalities in Higher Dimensions}\end{center}
For many years there have not been useful Bell inequalities for systems of higher dimensions. Only recently, a generalisation of the CHSH inequality to $d \otimes d$ dimensions was found and studied\cite{gen_chsh, gen_chsh2}. The corresponding quantity reads
\begin{equation}\begin{split} I_{d} = \sum_{k=0}^{\left\lfloor d/2 \right\rfloor -1} & \{[P(A_{1}=B_{1}+k)+P(B_{1}=A_{2}+k+1)+\\ & +P(A_{2}=B_{2}+k)+P(B_{2}=A_{1}+k)] -\\
& - [P(A_{1}=B_{1}-k-1)+P(B_{1}=A_{2}-k)+ \\
& + P(A_{2}=B_{2}-k-1) + P(B_{2}=A_{1}-k-1)]\}\end{split}\end{equation}
where $P(A_{i}=B_{j}+k)$ is the probability for the $i$-th measurement in the first system to yield the value of the $j$-th measurement in the second system plus $k$ modulo $d$ and $\left\lfloor . \right\rfloor$ is the floor-function (i.e. the function yielding the integer part of its argument). Due to the very general formulation in terms of computational basis vectors (which are always to be understood as modulo $d$), any of these inequalities can be applied to any $d' \otimes d'$-dimensional system, even if $d \neq d'$.
For $d = 2$ this quantity is bounded by 3 for LHV models. The corresponding inequality
\begin{equation} I_{2} = P(A_{1}=B_{1})+P(B_{1}=A_{2}+1)+P(A_{2}=B_{2})+P(B_{2}=A_{1}) \leq 3\end{equation}
is fully equivalent to the original CHSH inequality.\\
For any higher dimension $d > 2$, the expression is bounded by $I_{d} \leq 2$ for local realistic theories (while in all cases completely nonlocal theories could reach values up to $I_{d} = 4$).\\
A very central question is, if these ``generalised CHSH inequalities'' are optimal Bell inequalities in the same sense the original CHSH inequality is. As of yet, this question cannot be answered definitely, although there are certain unexpected properties of the higher dimensional inequalities, that suggest that they may not be. For example, the maximal violation of these higher dimensional Bell inequalities occurs for nonmaximally entangled states, i.e. a maximally entangled state yields a smaller violation than certain less entangled ones (the higher the dimension of the system, the bigger this difference gets)\cite{maxbell3viol, maxbell3viol2}.\\
This result offers two possible interpretations. Either the inequalities are in fact not optimal and there are others that do not show this property, or there is a more fundamental difference between quantum nonlocality and quantum entanglement than was known so far. While the latter case will be discussed later in reference to entanglement measures, the first case raises the further question of how to design better Bell inequalities. Using more general measurements, such as positive operator valued measurements (POVMs \cite{povm}) would lead to inequalities in which the maximal possible values could be attained even by separable states\cite{gisin_filters}, so it seems that a more appropriate way of generalising the CHSH inequality would be applying more than two von Neumann measurements per system.\\

\begin{center}\emph{Bell Operators}\end{center}
Often it is much easier to work not with the scalar Bell inequality itself, but to construct the corresponding Bell operator $B$ and formulate the inequality via this operator and the studied state, e.g.
\begin{equation}\label{bellopineq} \left\langle\Psi\right|B\left|\Psi\right\rangle \leq 2 \end{equation}
In the case of a two qubit system, the the single measurements correspond to observables of the form $\vec{a} \vec{\sigma} \otimes \vec{b} \vec{\sigma}$ (with $\left|\vec{a}\right| = \left|\vec{b}\right| = 1$), such that the CHSH inequality assumes the form (\ref{bellopineq}) with\cite{braunstein}
\begin{equation} B = \vec{a_{1}} \vec{\sigma} \otimes (\vec{b_{1}}+\vec{b_{2}}) \vec{\sigma} + \vec{a_{2}} \vec{\sigma} \otimes (\vec{b_{1}} - \vec{b_{2}}) \vec{\sigma} \end{equation}
while for higher dimensions, the observables have to be constructed via projectors instead of Pauli matrices (as in ref. \cite{maxbell3viol}).\\
Bell operators can easily be rewritten as nonoptimal entanglement witnesses $W$\cite{terhal, brussbw}
\begin{equation} W = 2\mathds{1} - B \end{equation}
since then the defining properties of an entanglement witness follow directly from the fact that all separable states are local.\\

\begin{center}\emph{Hidden Nonlocality}\end{center}
A somewhat puzzling feature of quantum nonlocality is that it can be ``hidden'' in states, such that the state itself is local (does not violate any Bell inequality), but local operations in the subsystems yield a nonlocal state\cite{hiddennl, gisin_filters}.\\
A very simple example is given by the two qubit state 
\begin{equation}\rho(\lambda,\alpha) = \lambda \left|\Psi_{\alpha}\right\rangle\left\langle\Psi_{\alpha}\right| + \frac{1-\lambda}{2}(\left|\uparrow\uparrow\right\rangle\left\langle\uparrow\uparrow\right| + \left|\downarrow\downarrow\right\rangle\left\langle\downarrow\downarrow\right|)\end{equation}
where $\left|\Psi_{\alpha}\right\rangle = \alpha\left|\uparrow\downarrow\right\rangle + \beta\left|\downarrow\uparrow\right\rangle$ and $\beta$ such that $\alpha^2+\beta^2=1$. This state is local if $\lambda \leq (1+\alpha^2\beta^2)^{-1}$.\\
If now the local polarisation dependant filters 
\begin{equation} F_{A} = \begin{pmatrix} \sqrt{\beta/\alpha} & 0 \\ 0 & 1 \end{pmatrix} \ \ \ \ \ \ \ F_{B} = \begin{pmatrix} 1 & 0 \\ 0 & \sqrt{\beta/\alpha} \end{pmatrix}\end{equation}
are applied to both subsystems accordingly, the resulting state is
\begin{equation} \rho'(\lambda, \alpha) = \frac{1}{N} \left[2\lambda\alpha\beta \left|\Psi^{-}\right\rangle\left\langle\Psi^{-}\right| + \frac{1-\lambda}{2}(\left|\uparrow\uparrow\right\rangle\left\langle\uparrow\uparrow\right|+\left|\downarrow\downarrow\right\rangle\left\langle\downarrow\downarrow\right|)\right]\end{equation}
where $N = 2\lambda\alpha\beta+(1-\lambda)$ is a normalisation-factor and $\left|\Psi^{-}\right\rangle$ is the Bell singlet state (\ref{bellstates}).\\
For special choice of $\lambda$ and $\alpha$ (for example $\lambda = 0.9$ and $\alpha\beta = 0.2$), the initial state $\rho$ is local, while the final state $\rho'$ is nonlocal. This result makes the identification of the term ``local'' with ``not violating any Bell inequality'' seem flawed, since nonlocality should not be creatable by local actions on any local state. However, there are a few open questions in this field which might lead to a different way to resolve this problem. Firstly, it is unknown whether local states containing such ``hidden nonlocality'' can be described by LHV models. Secondly, filtering and renormalising a state implies postselection (i.e. discarding or keeping systems from an ensemble according to the result of certain measurements), which can generally be used to enforce correlations (as in entanglement distillation). This means that one should not consider a single state conversion $\rho \rightarrow \rho'$, but a conversion of ensembles $\rho^{\otimes n} \rightarrow \rho'^{\otimes m}$ with $m \leq n$. Since $\rho$ being local does not involve $\rho^{\otimes n}$ being local\cite{peres_local}, the question should be asked whether there is a case in which $\rho^{\otimes n}$ is local and $\rho'^{\otimes m}$ is nonlocal (for appropriate $n$ and $m$). This however, turns out to be difficult, due to lack of trustworthy ways to determine if a high-dimensional state is local or nonlocal.\\

\begin{center}\emph{Bell Inequalities and Distillation}\end{center}
A priori, Bell inequalities do not show any special behaviour concerning distillable or undistillable states. It was conjectured, that any nonlocal quantum state should be distillable, this however turned out to be untrue, since there are examples where bound entangled states can violate certain Bell inequalities -- even maximally\cite{bebell}. Although this might be surprising at first, it is not too disturbing, since the used Bell inequality is not optimal for the used system (in the sense mentioned above) and also because even for optimal setups, mixed states are known to yield maximal violations due to the degeneracy of the corresponding Bell operator\cite{braunstein}.\\
There are, however, certain Bell inequalities that can only be violated by distillable states\cite{maxbell3viol}. Also, often the Bell-witnesses are decomposable into completely positive and completely co-positive maps as in eq. (\ref{mapdecomp}), such that they can only detect NPT states.\\

\begin{center}\emph{Bell Inequalities as a Measure for Entanglement}\end{center}
After all this discussion, it can be said that the magnitude of violation of Bell inequalities does not seem to pose a good measure for entanglement, for a number of reasons:
\begin{itemize}
	\item There are entangled states that do not violate any Bell inequality and that even admit LHV models\cite{werner_lhv}, some of which are even useful for teleportation\cite{popescu_bitele}, such that only a weaker form of the first desired property is fulfilled, namely: $\rho$ is separable $\Rightarrow E(\rho) = 0$, while the converse statement does not hold.
	\item The maximal violation can occur for nonmaximally entangled states, also there are mixed states that can reach this maximum. Hence this measure completely fails to identify maximally entangled states.
	\item Bell inequalities are not non-increasing under LOCC.
	\item Since no state can ever exceed the maximal value for any Bell inequality, the associated measure cannot be additive.
	\item For systems of higher dimension than qubits, the maximal violation of a Bell inequality for any given state is very hard to compute, since this task is a matter of multi-nonlinear numerical optimisation.
\end{itemize}

\subsection{Geometric Measures}
A whole class of geometrically highly intuitive measures for entanglement can be formulated via geometric distances\cite{vedral_plenio_1}. These are a quantitative generalisation of entanglement witnesses in the sense that they are based on the convexity of the set of separable states, from which follows that there is exactly one closest separable state to each entangled state (in any metric, on which this state may depend). The idea is that this nearest separable state should reproduce the classical portion of the correlations contained in an entangled state, so that any remaining difference between the two states is solely due to quantum entanglement.\\
This measure is simply defined by
\begin{equation} E(\rho) = \min_{\sigma \in S} D(\rho, \sigma) \end{equation}
where $D(\rho, \sigma)$ is any measure of distance between the two density matrices $\rho$ and $\sigma$ and the minimum is taken over all separable states $\sigma$. Note that $D(\rho, \sigma)$ does not necessarily have to be a real metric (as will be seen in \ref{relentropy}).\\
Obviously, whether this represents a good measure for entanglement depends strongly on the choice of $D(\rho, \sigma)$. It was shown\cite{vedral_plenio_2} that such a measure is invariant under unitary transformations, nonincreasing under LOCC and yields its minimal value of $E = 0$ for separable states and for separable states only, if the used distance measure satisfies the following criteria:
\begin{enumerate}
	\item $D(\rho, \sigma) \geq 0$, where equality holds iff $\rho = \sigma$
	\item $D(\rho, \sigma) = D(U \rho U^{\dagger}, U \sigma U^{\dagger})$ for any unitary operation $U$
	\item $D(\mathrm{Tr}_{X}\rho, \mathrm{Tr}_{X}\sigma) \leq D(\rho, \sigma)$ where $\mathrm{Tr}_{X}$ is a partial trace
	\item $\sum_{i} p_{i} D(\rho_{i} / p_{i} , \sigma_{i} / q_{i}) \leq \sum_{i} D(\rho_{i}, \sigma_{i})$ where $\rho_{i} = V_{i} \rho V_{i}^{\dagger}$, $\sigma_{i} = V_{i} \sigma V_{i}^{\dagger}$, the $\{V_{i}\}$ represent a POVM (i.e. are positive operators satisfying $\sum_{i} V_{i} V_{i}^{\dagger} = \mathds{1}$) and $p_{i} = \mathrm{Tr}(\rho_{i})$ and $q_{i} = \mathrm{Tr}(\sigma_{i})$
	\item $D(\sum_{i} P_{i} \rho P_{i}, \sum_{i} P_{i} \sigma P_{i}) = \sum_{i} D(P_{i} \rho P_{i}, P_{i} \sigma P_{i})$ where the $\{P_{i}\}$ are any set of orthogonal projectors
	\item $D(\rho \otimes P, \sigma \otimes P) = D(\rho, \sigma)$ for any projector P.
\end{enumerate}
Obviously, condition (1) ensures $E(\rho) = 0$ iff $\rho$ is separable and condition (2) ensures unitary invariance. Less trivially, conditions (2)-(6) are sufficient for the measure to be nonincreasing under LOCC.\\

\subsubsection{Hilbert-Schmidt-Distance\label{hilbertschmidt}}
The probably most obvious choice for $D(\rho, \sigma)$ is the standard Hilbert-Schmidt metric
\begin{equation} D(\rho, \sigma) = \left\|\rho - \sigma\right\| = \sqrt{\mathrm{Tr}(\rho - \sigma)^2} \end{equation}
Although the conditions (3) - (6) have not been proven (or falsified) for this metric\cite{vedral_plenio_2}, it shows some remarkable features and is therefore one of the commonly used distance measures in entanglement quantification.\\

\begin{center}\emph{The Bertlmann-Narnhofer-Thirring theorem}\end{center}
The analogy between geometric entanglement measures and entanglement witnesses is most significant in the Hilbert-Schmidt metric. Restricting the set of entanglement witnesses to normalised operators, i.e. operators satisfying $\left\|A-\alpha\mathds{1}\right\|_{2} = 1$ (where $\alpha = \mathrm{Tr}(A)/d$), the maximal value of the witness-inequality (\ref{ew0}) for a given state
\begin{equation} B(\rho) = \max_{A}\left[\min_{\sigma \in S} \left\langle\sigma | A\right\rangle - \left\langle \rho | A\right\rangle \right] \end{equation}
(where the maximum is taken over all normalised entanglement witnesses $A$ and the minimum is taken over all separable states $\sigma$) equals the Hilbert-Schmidt distance $E_{HS}(\rho)$ of this state to the set of separable states\cite{bertl_narn_thirring}:
\begin{equation} B(\rho) = E_{HS}(\rho) \ \ \ \forall \ \rho \end{equation}
This is known as the Bertlmann-Narnhofer-Thirring theorem.\\
Since $B(\rho)$ is defined as a maximum and $E_{HS}(\rho)$ is defined as a minimum, upper and lower bounds are very easily obtained:
\begin{equation} \min_{\sigma \in S}\left\langle\sigma - \rho |\frac{\omega - \rho}{\left\|\omega-\rho\right\|} \right\rangle \leq B(\rho) = E_{HS}(\rho) \leq \left\|\omega-\rho\right\| \ \ \ \forall \ \omega \end{equation}
where the lower bound is attained for a geometric entanglement witness in which the minimum is taken over all separable states $\sigma$.\\

\begin{center}\emph{Finding the nearest separable state}\end{center}
Similar to other distance measures, the nearest separable state to any given entangled state is in general rather difficult to find in the Hilbert-Schmidt metric. However, there is a way to check if a ``guessed'' state happens to be the nearest separable state\cite{bertl_oew}.\\
Namely, a separable state $\tilde{\sigma}$ is the closest separable state to a given entangled state $\rho$, iff the operator
\begin{equation} C = \frac{\tilde{\sigma}-\rho-\left\langle\tilde{\sigma} | \tilde{\sigma}-\rho\right\rangle\mathds{1}}{\left\|\tilde{\sigma}-\rho\right\|}\end{equation}
is an entanglement witness.\\
This follows from the construction of $C$, which is an (optimal) geometric entanglement witness if and only if $\tilde{\sigma}$ really is the nearest separable state to $\rho$, since otherwise the hyperplane spanned by $C$ would intersect the set of separable states and hence $C$ would not be an entanglement witness.\\

\begin{center}\emph{Finding the nearest PPT state}\end{center}
Even in cases with more complex or nonintuitive geometry, such that the nearest separable state can not be guessed, there is a very operational method for finding the closest PPT state (which in many cases is also the closest separable state, and else gives a good lower bound on the distance) by means of Lagrangean methods\cite{verstraete}.\\
\begin{enumerate}
	\item First, one has to find the eigenvalue decomposition of the partially transposed density matrix of the entangled state $\rho$
	\begin{equation} \rho^{T_{A}} = U D U^{\dagger} \end{equation}
where $D$ is a diagonal matrix and $U$ is a unitary transformation.
	\item One then defines the unique positive semidefinite normalised diagonal matrix $E$ with diagonal entries 
	\begin{equation} e_{i} = \max(d_{i} + \lambda, 0) \end{equation}
	where $d_{i}$ are the elements of $D$ and the value of $\lambda$ is defined by the normalisation $\mathrm{Tr} E = 1$.
	\item The nearest PPT-state to $\rho$ is obtained as
	\begin{equation} \tilde{\sigma} = (U E U^{\dagger})^{T_{A}} \end{equation}
\end{enumerate}
Now, if $\tilde{\sigma}$ is in fact a state (i.e. $\tilde{\sigma} \geq 0$), then it is indeed the nearest PPT state to $\rho$. However, there are special cases in which the outcome yielded by this procedure does not happen to be positive. Even in these cases $\left\|\tilde{\sigma}-\rho\right\|$ is a useful lower bound on the actual distance.\\

\subsubsection{Quantum Relative Entropy\label{relentropy}}
The quantum relative entropy of two states (also called (relative) entropy of entanglement, when used as an entanglement measure) is a measure for the overlap of two density matrices (in the sense of statistical distinguishability\cite{vedral_plenio_j_k}) defined by
\begin{equation} D(\rho, \sigma) = \mathrm{Tr}(\rho \log \rho - \rho \log \sigma) \end{equation}
Obviously, the quantum relative entropy is not a metric (as it does not satisfy the triangle inequality and is not symmetric in $\rho$ and $\sigma$), however, the entropy of entanglement $E_{RE}$ is a remarkable measure for entanglement, as it is closely related to the entanglement of formation $E_{F}$ and the entanglement of distillation $E_{D}$\cite{henderson}, as will be discussed in section \ref{entofdistill}. Also, for pure states $E_{RE}$ reduces to the von Neumann entropy of the subsystems, i.e. the entanglement measure for pure states (\ref{puremeasure})\cite{vedral_plenio_2}.\\
Explicit analytical computation of the quantum relative entropy is only possible in special highly symmetric cases\cite{measuresymmetry}, in which examples can be found that prove the measure to be nonadditive.\\

\subsubsection{Bures-Distance}
The last distance measure that shall be discussed here is the Bures-metric\cite{vedral_plenio_2}
\begin{equation} D(\rho, \sigma) = 2 - 2\sqrt{F(\rho, \sigma)} \end{equation}
\[ F(\rho, \sigma) = \left[\mathrm{Tr}(\sqrt{\sigma}\rho\sqrt{\sigma})^{\frac{1}{2}}\right]^2 \]
Like the quantum relative entropy, the Bures-metric satisfies all of the conditions discussed for geometric entanglement measures. It is closely related to the question of experimentally distinguishing one state from another, as in the field of unambiguous state discrimination\cite{usd}.\\
The Bures-distance can also be seen in context with LOCC, as $F$ can also be written as\cite{fuchs_caves}
\begin{equation} F(\rho, \sigma) = \min_{\{A_{i}\}} \sum_{i}\sqrt{\mathrm{Tr}(A_{i} \rho A_{i}^{\dagger})} \sqrt{\mathrm{Tr}(A_{i} \sigma A_{i}^{\dagger})} \end{equation}
where the minimum is taken over all sets of positive operators $\{A_{i}\}$ satisfying $\sum A_{i}^{\dagger}A_{i} = \mathds{1}$, i.e. POVMs.\\

\subsection{Entanglement of Formation and Entanglement of Distillation\label{entofdistill}}
Historically, the entanglement of formation $E_{F}$ and the entanglement of distillation $E_{D}$ were the first attempts to quantify quantum entanglement. Until today, they are the only measures for entanglement with a direct physical interpretation, making them incomparably important and meaningful measures and give them a unique perspective.\\

\begin{center}\emph{Entanglement of Distillation}\end{center}
The entanglement of distillation of a state $\rho$ is defined as the maximal asymptotic (in the limit of high numbers of particles) yield of distilled maximally entangled states per copy of the input state $\rho$, i.e.\cite{qecc}.
\begin{equation} E_{D} = \max \ \lim _{n \rightarrow \infty} \frac{m(n)}{n} \end{equation}
where $n$ is the number of input states $\rho$, $m$ is the number of maximally entangled output states and the maximum is taken over all possible distillation protocols.\\
By definition, this measure obviously is normalised to $0 \leq E_{D} \leq 1$ and the upper bound is reached iff $\rho$ is maximally entangled. Furthermore, the entanglement of distillation satisfies $E_{D} = 0$ for all separable states, but not for separable states only, since bound entangled states cannot be distilled and hence have $E_{D} = 0$ without being separable.\\
Since very little is known about distillation protocols apart from the few explicit examples that have been established, the entanglement of distillation is hardly computable. However, since it is defined as a maximum, lower bounds can easily be obtained by applying any particular protocol. Also, the entanglement of formation $E_{F}$ (which will be discussed subsequently) always gives an upper bound on the entanglement of distillation.\\
Since there are distillation protocols that use only one-way communication, these define another form of entanglement of distillation $E_{D_{1}}$, which obviously cannot exceed the general entanglement of distillation, but is proven to be lower for certain states (there even are examples of states for which $E_{D_{1}} = 0 < E_{D}$), such that
\begin{equation} E_{D_{1}} \leq E_{D} \leq E_{F} \end{equation}
Also, when dealing with more complicated situations (e.g. more than two parties), the one-way entanglement of distillation can be nonsymmetric, i.e. can depend on the direction of the allowed communication: $E_{D_{1}}^{A \rightarrow B} \neq E_{D_{1}}^{B \rightarrow A}$.\\
Due to phenomenae like the bound entanglement activation effect (as discussed in (\ref{activateboundent})) and the probable existence of NPT bound entanglement, the entanglement of distillation seems to be nonadditive\cite{horo_limits}.\\

\begin{center}\emph{Entanglement of Formation}\end{center}
The entanglement of formation of a state $\rho$ is defined as the optimal asymptotic (in the limit of high particle numbers) conversion yield of the number of maximally entangled input states needed per output state $\rho$\cite{qecc}. In other words, it is the answer to the question: How many maximally entangled states are necessary, to form one copy of a mixed-entangled state $\rho$, i.e.
\begin{equation} E_{F} = \min \ \lim _{m \rightarrow \infty} \frac{m}{n(m)} \end{equation}
where $m$ is the number of maximally entangled input states, $n$ is the number of output states $\rho$ and the minimum is taken over all means of preparing $\rho$ from maximally entangled states.\\
Since the preparation entanglement cost for a pure state is given by the von Neumann entropy of one of its reduced density matrices (i.e. the entanglement measure for pure states (\ref{puremeasure})), it seems reasonable that the appropriate convex roof equals the entanglement of formation for a mixed state. Although on second thought it seems possible that there are more efficient ways of preparing an entangled state than to prepare each pure state in any optimal decomposition and mixing them, this was proven not to be the case\cite{entcost}, such that
\begin{equation} E_{F} = \min_{\{p_{i}, \left|\Psi_{i}\right\rangle\}} \sum_{i} p_{i} S(\rho_{i}) \label{eof}\end{equation}
where the minimum is taken over all decompositions $\{p_{i}, \left|\Psi_{i}\right\rangle\}$ satisfying $\rho = \sum_{i} p_{i} \left|\Psi_{i}\right\rangle\left\langle\Psi_{i}\right|$ and the $\rho_{i}$ are the reduced density matrices of the states $\left|\Psi_{i}\right\rangle$.\\
The entanglement of formation satisfies all of the desired conditions discussed above, except additivity, which has not been proven as of yet.\\
While it may seem surprising at first, that the entanglement of distillation does in general not equal the entanglement of formation (while they are both measures for the entanglement contained in a quantum state), it becomes obvious when considering special cases, such as bound entanglement, which by definition has $E_{D} = 0 < E_{F}$.\\
Unfortunately, the entanglement of formation is in general not easily computable (although unlike the entanglement of distillation, it can be computed numerically from eq. (\ref{eof}) in principle), but there are special cases in which a closed form can be given via the so called concurrence.\\

\begin{center}\emph{Concurrence}\end{center}
The entanglement of formation can be computed analytically in several special cases via a quantity called the concurrence $C$: $E_{F} = E_{F}(C)$, where $E_{F}$ ranges from 0 to 1 for $C$ going through the values from 0 to 1 as well, making the concurrence a suitable measure for entanglement itself\cite{2x2conc}.\\
In this way, $E_{F}$ can be computed for example for all states of two qubit systems:
\begin{equation} E_{F} = h\left(\frac{1+\sqrt{1-C^2}}{2}\right) \end{equation}
where $h(x)$ is the binary entropy function
\begin{equation} h\left(x\right) = -x \log x - (1-x) \log (1-x) \end{equation}
and the concurrence $C$ is given by
\begin{equation} C = \max(0, \lambda_{1} - \lambda_{2} - \lambda_{3} - \lambda_{4}) \end{equation}
with $\lambda_{i}$ being the eigenvalues of the matrix
\[ R = \sqrt{\sqrt{\rho}\tilde{\rho}\sqrt{\rho}} \]
\begin{equation} \tilde{\rho} = \left(\sigma_{y}\otimes\sigma_{y}\right) \rho* \left(\sigma_{y}\otimes\sigma_{y}\right) \end{equation}
\[ \sigma_{y} = \begin{pmatrix} 0 & -i \\ i & 0 \end{pmatrix} \]
in decreasing order and the asterisk denoting complex conjugation.\\
This concurrence can be generalised to higher dimensions\cite{audenaert}
\begin{equation} C = \min_{\{p_{i}, \left|\Psi_{i}\right\rangle\}} \sum_{i} p_{i} \max_{\{U_{i}, V_{i}\}} \left|\left\langle\Psi_{i}\right|\left(U_{i} \otimes V_{i}\right)^{T} S \left(U_{i} \otimes V_{i}\right)\left|\Psi_{i}\right\rangle\right| \end{equation}
where the minimum is taken over all ensembles $\{p_{i}, \left|\Psi_{i}\right\rangle\}$ realising $\rho$ and the maximum is taken over all $U_{i} \in SU(d_{1})$ and $V_{i} \in SU(d_{2})$, with $d_{1, 2}$ being the dimensions of the two subsystems, and where
\begin{equation} S = \left(\sigma_{y} \oplus 0_{d_{1}-2}\right)\otimes\left(\sigma_{y} \oplus 0_{d_{2}-2}\right) \end{equation}
is a generalised version of the spin-flip operator $\sigma_{y}\otimes\sigma_{y}$ from the $2\otimes2$-dimensional case. This generalisation however seems rather worthless, as it is exactly as difficult to compute as the entanglement of formation (\ref{eof}) itself.\\
There are several classes of states for which a closed form of $C$ (or $E_{F}$ itself) has been found (for example the isotropic state in arbitrary dimensions\cite{conciso}), as well as multiple bounds (e.g. \cite{minert}) and methods to simplify the variational problem (\ref{eof}), reducing it for example to a variational problem over finite-sized matrices\cite{audenaert}. For general states however, there still exists no closed form of computation for the entanglement of formation.\\

\begin{center}\emph{Limits for entanglement measures}\end{center}
Due to their direct physical interpretation, the entanglement of formation and the entanglement of distillation pose new restrictions on other entanglement measures. It is more than reasonable to demand
\begin{equation} E_{D} \leq E \leq E_{F} \label{limitem}\end{equation}
for any entanglement measure $E$, as a state cannot contain more entanglement than is needed to create it and can neither contain less entanglement than can be distilled from it. In the light of this it seems appropriate not to require entanglement measures to satisfy conditions that ``seem'' resonable (especially since quantum mechanics has proven to behave contra-intuitively on several other occasions), but to relate the demands to objectively good entanglement measures, which $E_{D}$ and $E_{F}$ are, regardless of which of the discussed conditions are satisfied.\\
For pure states, both equalities in (\ref{limitem}) hold\cite{henderson}, such that the entanglement measure for pure states is uniquely given as discussed in \ref{puremeasure}. For mixed states however, the problem is more difficult. It seems, that a single quantity does not suffice to describe the entanglement contained in a quantum state (this is already suggested by the entanglement of formation not being equal to the entanglement of distillation).\\
An entanglement measure $E$ was proven to satisfy (\ref{limitem}), if it is additive, convex and reduces to the von Neumann entropy of the reduced density matrices for pure states\cite{horo_limits}. The only entanglement measure known to satisfy condition (\ref{limitem}) apart from $E_{F}$ and $E_{D}$ themselves, is the entropy of entanglement, i.e. the geometric entanglement measure induced by the quantum relative entropy (as discussed in \ref{relentropy}).\\

\subsection{Schmidt Numbers\label{schmidtnum}}
The concept of Schmidt ranks for pure states can be generalised to mixed states\cite{schmidt}. The Schmidt number of a given density matrix $\rho$ is defined as the maximal Schmidt rank, that is at least necessary to construct it, i.e. the number $k(\rho)$ such that
\begin{itemize}
	\item There is no decomposition of $\rho$ into pure states which all have Schmidt rank less than $k(\rho)$
	\item There is a decomposition of $\rho$ into pure states which all have a Schmidt rank of at most $k(\rho)$
\end{itemize}
The Schmidt number of a state determines how many degrees of freedom are entangled within this state. Evidently, the lower bound $k(\rho) = 1$ is equivalent to $\rho$ being separable, while the upper bound $k(\rho) = d_{min} = \min(d_{1}, d_{1})$ indicates all present degrees of freedom being entangled.\\
For formal reasons, one might define $E = k(\rho) - 1$ as an entanglement measure, such that $E = 0$ iff $\rho$ is separable. The Schmidt number of a state is nonincreasing under LOCC, convex (in the sense that $\lambda k(\rho_{1}) + (1-\lambda) k(\rho_{2}) \leq k(\lambda \rho_{1} + (1-\lambda) \rho_{2})$) and unitary invariant, it is however not additive.\\
From the first of these properties follow strong limits on interconvertibility of states. If for example $k(\rho^{\otimes n}) < k(\omega^{\otimes n})$ for a certain number $n$, this means that $\rho$ cannot be converted to $\omega$ on a 1-to-1 basis, even if $k(\rho) = k(\omega)$. The behaviour of Schmidt numbers for high numbers of copies of a state can be very different, depending on the state -- in some cases, they grow to very large numbers, while in others they do not grow at all, such that $k(\rho) = k(\rho^{\otimes n})$.\\

\subsection{Robustness of Entanglement\label{robustness}}
\begin{figure}[ht!]\centering\includegraphics[width=80mm]{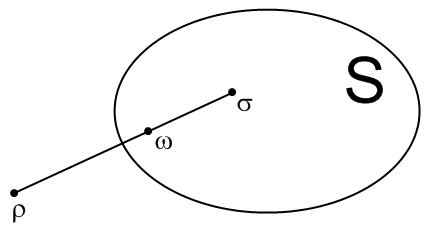}\caption[Visualisation of the robustness of entanglement]{Visualisation of $R(\rho || \sigma)$ (eq. (\ref{eq_robust})). The minimum is achieved when $\omega \in \partial S$. Note that the robustness measure is independant of any choice of metric.}\label{fig_robust}\end{figure}

The robustness of entanglement\cite{robustness} $R$ originated from the already discussed idea of describing mixed state entanglement not with a single measure, but with several parameters, in which context the robustness seems to be useful. It is a measure for how much a given quantum state $\rho$ has to be mixed with a separable state $\sigma$, so that the resulting state is separable and thus all entanglement is ``erased'':
\begin{equation} R\left(\rho || \sigma \right) = \min s\label{eq_robust}\end{equation}
where the minimum is taken over $s$ such that
\begin{equation} \omega\left(s\right) = \frac{1}{1+s}\left(\rho + s \sigma\right) \in S \end{equation}
There are two different choices of $\sigma$ that induce useful measures for entanglement. Firstly, the absolute robustness, where $R\left(\rho || \sigma \right)$ is minimised over all separable states $\sigma$:
\begin{equation} R(\rho) = \min_{\sigma \in S} R\left(\rho || \sigma \right) \end{equation}
and secondly, the random robustness, where $\sigma = 1/d \ \mathds{1}$ is the maximally mixed state.\\
Note that both these measures are well defined (since a neighborhood of the maximally mixed state is always separable and the geometry shown in Fig. \ref{fig_robust} can always be achieved). Furthermore, both measures vanish iff $\rho$ is separable.\\
The absolute robustness was shown to satisfy most of the discussed conditions, although it only is additive in the very weak sense that
\begin{equation} R(\rho \otimes \sigma) = R(\rho) \ \forall \ \sigma \ \in \ S \end{equation}
The computation of $R(\rho)$ is simplified significantly by the fact, that $R\left(\rho || \sigma \right)$ is a convex function of $\sigma$, meaning that any local minimum is also a global minimum, such that the optimisation over $\sigma$ can be done (numerically) in most cases. Hence, both the absolute and the random robustness of entanglement can be computed, given a means to faithfully detect entanglement in the given system. In several cases, they can even be calculated analytically. Surprisingly, the absolute robustness of entanglement turns out to be a quantification of the cross-norm criterion, as it can be written as\cite{further_ccn}
\begin{equation} R(\rho) = \left\|\rho\right\|_{\gamma} - 1\end{equation}
Also, several bounds on the robustness are known\cite{robustness}.\\

\subsection{Fidelity}
The mathematically most simple and straightforward measure for entanglement might be the fidelity. It is defined as the conversion probability from a given state $\rho$ to the set of maximally entangled states (which can be chosen to be real)
\begin{equation} F = \max_{\left|\Psi\right\rangle} \left\langle\Psi\right|\rho\left|\Psi\right\rangle \end{equation}
where the maximum is taken over all maximally entangled states $\left|\Psi\right\rangle$.\\
Although it is very useful to characterise quantum states (for example as a parameter for a family of states, such as in the already discussed Werner state (\ref{werner})), it is of rather little use in entanglement quantification, since is not nonincreasing under LOCC and is also nonzero for separable states. Thus, it cannot even be used to detect entanglement (except in special cases of families of states that are known to be separable/entangled for certain fidelities).\\

\subsection{Negativity}
Another direct quantification of an entanglement detection procedure is the so called negativity $N$. It is a simple quantification of the nonpositivity of the partially transposed density matrix of a state\cite{negativity}
\begin{equation} N = \frac{\left\|\rho^{T_{A}}\right\|_{1} - 1}{2} \end{equation}
such that $N = 0$ for all PPT states and $N = 1$ for maximally entangled states.\\
This measure seems very useful, as it is very easily computable. Nevertheless, it vanishes for many entangled states and is only applicable for bipartite systems and thus in general is not a good measure\begin{figure}[ht!]\centering\includegraphics[width=93mm]{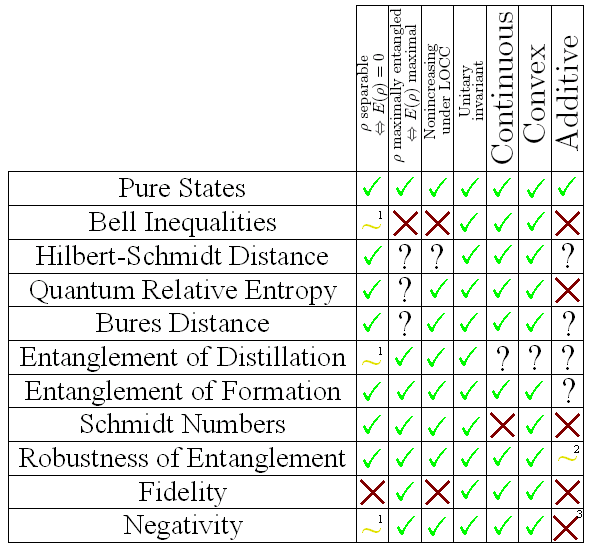}\caption[List of the discussed measures for entanglement and their properties]{In the light of results on various entanglement measures, espcially on the entanglement of formation and the entanglement of distillation, it seems as if the discussed properties of entanglement measures are not as fundamental as they were considered to be at first. Nevertheless, they offer a good overview of the usefulness of the discussed measures. A green checkmark stands for the property being satisfied by the entanglement measure, while a red 'x' represents the measure failing to do so. Question marks indicate properties that have not been proven one way or the other. Waves symbolise a property that is only satisfied in a weaker form.\newline
$^{1}: \rho$ is separable $\Rightarrow$ $E(\rho) = 0$, but not vice versa.\newline
$^{2}:$ Only a weak form of additivity is satisfied, as discussed in \ref{robustness}.\newline
$^{3}:$ Since the negativity is only defined for bipartite systems, this property makes no sense.}\label{fig:measures}\end{figure}.\\

\newpage
\section{Hilbert Space Geometry and Examples}
One of the ultimate goals in quantum information theory is to completely characterise and fully understand the geometry of multipartite Hilbert spaces and the properties of quantum states therein.\\

\subsection{Unipartite Systems}
Unipartite Hilbert spaces can be visualised by the concept of Bloch vectors\cite{bloch}. A $d \otimes d$-dimensional density matrix $\rho$ can always decomposed in the form
\begin{equation} \rho = \frac{1}{d}\mathds{1} + \vec{b}\vec{\Gamma} \label{eq_bloch}\end{equation}
where $\vec{\Gamma}$ is a list of $(d^2-1)$ mutually orthogonal traceless matrices, which -- together with the identity matrix -- form a basis of the Hilbert-Schmidt space and $\vec{b}$ is a list of $(d^2-1)$ coefficients, the so called Bloch vector, such that $\vec{b}\vec{\Gamma}$ is a linear combination of all the basis matrices. Obviously, the state space of $d \otimes d$-dimensional density matrices has $d^2-1$ dimensions, since the coefficient of the identity matrix is given by $\mathrm{Tr} \rho = 1$. Note that this map is not bijective, as for any choice of basis $\vec{\Gamma}$ any given density matrix $\rho$ corresponds to a certain Bloch vector $\vec{b}$, but not every given $\vec{b}$ results in a density matrix, as the decomposition (\ref{eq_bloch}) does not necessarily provide positivity.\\
For a single qubit, i.e. a $2\times2$-dimensional density matrix, the state space is very simple. The commonly used matrix basis consists of the three Pauli matrices (\ref{pauli}) and the decomposition (\ref{eq_bloch}) assumes the form
\begin{equation} \rho = \frac{1}{4}\left(\mathds{1} + \vec{a}\vec{\sigma}\right) \end{equation}
with $\left|\vec{a}\right| \leq 1$, where equality holds iff $\rho$ is a pure state and $\left|\vec{a}\right|$ gets smaller the more mixed $\rho$ is.\\
Already for $3\times3$-dimensional density matrices, i.e. states of a single qutrit, the Bloch sphere becomes more complicated and the choice of basis less obvious. The most commonly used basis matrices are the Gell-Mann matrices (\ref{ggm}) and the Weyl operators (\ref{weyl}), which each result in a certain allowed interval for $|\vec{b}|$ where all density matrices are found. However, unlike in the qubit case, there are Bloch vectors that do not correspond to states. Thus, the state space for a single qutrit is already rather complicated.\\

\subsection{Bipartite Systems}
Since the lowest-dimesional bipartite system (the system of two qubits, which is $(2\times2)^2-1=15$ dimensional) has already significantly more dimensions than a system of one qutrit ($3^2-1 = 8$ dimensions), which has already been found to be rather nontrivial, bipartite systems are generally more difficult to analyse. Furthermore, in bipartite systems the concept of entanglement starts to play an important role, such that there are more properties of states that need to be taken into account and implemented into the picture.\\
In most cases, it is very difficult to understand the underlying geometry by studying mathematical properties. This is why often only 2- or 3-dimensional simplices (i.e. subspaces) are studied, since these can be visualised and help to gain an intuitive and geometrical understanding.\\

\subsubsection{Classification of States}
\begin{figure}[ht!]\centering\includegraphics[width=120mm]{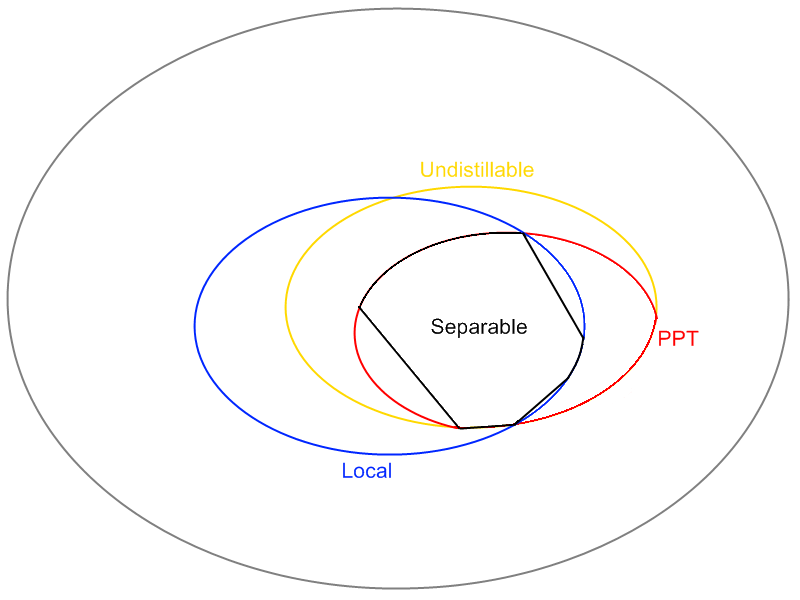}\caption{Schematic picture of the different classes of states}\label{fig_geometry}\end{figure}

In the earlier sections of this work, a basic understanding of the different types of bipartite quantum states has been established. These types of quantum states shall now be reviewed and discussed in further detail.\\
There are four basic criteria (which are relevant in the context of quantum information theory) by which bipartite states can be categorised. Firstly, they can be either separable or entangled. Secondly, they can be either distillable (free entangled) or undistillable (separable or bound entangled). Thirdly, they can be either local or nonlocal (i.e. satisfy all Bell inequalities or violate some of them). Finally, they can be either PPT or NPT (i.e. remain positive under partial transposition or not).\\
While in each of these four distinctions a given quantum state has exactly one of the possible properties (e.g. it is either PPT or NPT, but not both and not neither), most combinations of these properties with properties from other categories are possible (such as 'PPT and entangled' or 'distillable and local'), as visualised in Fig. \ref{fig_geometry}.\\

\begin{center}\emph{Separable and entangled states}\end{center}
The most important distinction in this context is the one between separable and entangled states. As was already thoroughly discussed throughout this work, this distinction is highly nontrivial, as the definition of separable states
\begin{equation} \rho = \sum_{i} p_{i} \left|\Psi_{i}^{A}\right\rangle\left\langle\Psi_{i}^{A}\right| \otimes \left|\Psi_{i}^{B}\right\rangle\left\langle\Psi_{i}^{B}\right| \end{equation}
(where the $\{p_{i}\}$ are probabilities) is in general very difficult to check and there exists no closed definition for entangled states apart from not being separable.\\
Still, some properties of the set of separable states $S$ and the set of entangled states $S^{C}$ are known\cite{bertl_narn_thirring}:
\begin{itemize}
	\item $S$ is a convex set (consequently, $S^{C}$ is not).
	\item $\mathrm{dim} S = \mathrm{dim} S^{C} = D^2 - 1$, where $D = d_{1} \ d_{2}$ is the dimension of the composite system. This means that both $S$ and $S^{C}$ are thick everywhere on $\mathcal{H}$ and not sets of measure zero.
	\item All separable density matrices with rank one or two (i.e. pure states and mixtures of no more than two pure states) are located on the border of $S$, while density matrices of higher rank (i.e. mixtures of more than two pure states) form the interior of $S$, but can also be found on the border.
	\item Whereever there is a mixed state of $n$ pure states on the border of $S$, there is an at least $n$-dimensional face of $S$. If the dimensions of the two subsystems are equal ($d_{1} = d_{2} = d$), then there exist at least $d^2$-dimensional faces.
	\item $S$ is invariant under transformations of the form $\Lambda^{A}\otimes\Lambda^{B}$, where $\Lambda^{i}$ are positive maps (as discussed in \ref{pms}).
\end{itemize} \vspace{5mm}

\begin{center}\emph{Distillable and undistillable states}\end{center}
Since very little is known about distillation and distillation protocols, it is as of yet not possible to completely define the set of distillable states or the set of undistillable states for general systems. Nevertheless, in big parts of the state space the question can be answered, for example by means of the reduction criterion or the PPT-criterion (since violation of the reduction criterion implies distillability, while satisfaction of the PPT-criterion implies undistillability, as discussed in \ref{pms}). Pure entangled states however are always distillable.\\
Since very little is known about the geometry of bound entanglement, it is still unknown if the set of undistillable states is convex.\\

\begin{center}\emph{Local and nonlocal states}\end{center}
Like distillability, nonlocality of a state (in the sense of violating Bell inequalities) is in general not completely characterised, as Bell inequalities for higher dimensions than $2\otimes2$ are merely beginning to be understood. While the $2\otimes2$ case is solved (as discussed in \ref{bell}), for higher dimensions there are even less possibilities to answer this question for a given state as there are for distillability. Apart from all separable states being local and pure entangled states being nonlocal, there is no way to tell if a state is local, while nonlocality can only be proven by explicitly finding a Bell inequality that is violated by this state.\\
Since Bell inequalities are linear functions of states (since they can be written as expectation values of Bell operators), a convex sum of local states remains local. The set of local states is therefore a convex one.\\

\begin{center}\emph{PPT and NPT states}\end{center}
The question of positivity under partial transposition of a state lacks a direct physical meaning, nevertheless it is a very powerful tool, since it is very easily computable for all kinds of multipartite quantum states and offers a great deal of information on separability and distillability of a state. If a state is separable, it has to be PPT and a PPT state hast to be undistillable. Conversely, a distillable state has to be NPT and an NPT state has to be entangled.\\

\begin{figure}[ht!]\centering\includegraphics[width=80mm]{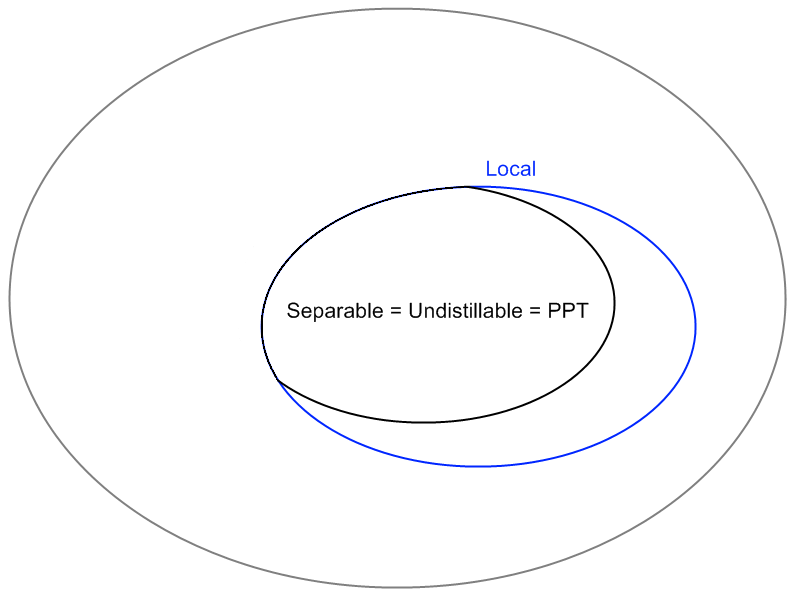}\caption{Schematic picture of the state space in $2\otimes2$ and $2\otimes3$ dimensions}\label{fig_2x2+3}\end{figure}

\begin{center}\emph{Low dimensional cases}\end{center}
In certain lower dimensions, the geometry of the state space is much more simple and some of the four criteria become equivalent.\\
In $2\otimes d_{2}$ dimensions (with $d_{2}$ being an arbitrary integer), nonpositivity under partial transposition is necessary and sufficient for distillability, i.e. the set of NPT states is the same as the set of distillable states (there is no NPT bound entanglement)\cite{dclb}.\\
In $2\otimes2$ and $2\otimes3$ dimensions, positivity under partial transposition is necessary and sufficient for separability\cite{horo_pms}, such that there is no PPT bound entanglement either. Consequently, in these cases there is no bound entanglement at all, PPT (NPT) is equivalent to separability (entanglement) and undistillability (distillability)\cite{horo_2x2distill}. Therefore, for these systems there are only two distinct criteria left (separability and locality) and the geometry is considerably simplified (see Fig. \ref{fig_2x2+3}).\\

\begin{center}\emph{Schmidt classes}\end{center}
Apart from the classification above, density matrices can also be characterised in terms of their Schmidt numbers (as discussed in \ref{schmidtnum}). The Schmidt class $S_{k}$ is defined as the set of all states, whose Schmidt number does not exceed $k$, i.e. all states $\rho$ satisfying $k(\rho) \leq k$.\\
Due to the convexity of Schmidt numbers, each Schmidt class forms a convex subset of the set of all states, such that
\begin{equation} S_{1} \subset S_{2} \subset ... \subset S_{d_{min}} \end{equation}
$S_{1} = S$ is the set of separable states and $S_{d_{min}}$ is the complete set of states, apart from this however, there are hardly any known links between Schmidt classes and the other sets of states discussed above. Evidence has been found\cite{schmidtbe} that (at least in special cases) bound entangled states do not seem to have maximal Schmidt number, but further study in this area is needed.\\
Due to their convexity, Schmidt classes can be described by appropriate witness operators -- so called Schmidt witnesses\cite{schmidtwitness}, that separate a certain $S_{n}$ from all states $\rho \notin S_{n}$.\\

\subsubsection{Systems of two QuBits}
Systems of two qubits and systems of one qubit and one qutrit are the only completely solved ones. Since systems consisting of equal subsystems are much more practical in use, the $2\otimes2$ case is of much more interest than the $2\otimes3$ one.\\
Since the PPT criterion is necessary and sufficient for separability in this case, it is very simple to find out if any given state is separable (and thus undistillable) or entangled (and distillable, as there is no bound entanglement in this case). The question of locality or nonlocality can be easily answered by means of the criterion (\ref{2x2bell}).\\

\begin{center}\emph{Bloch decomposition for two qubits}\end{center}
As was already mentioned in eq. (\ref{bloch2x2}), any density matrix of a two qubit system can be written as
\begin{equation} \rho = \frac{1}{4}(\mathds{1} + \vec{a} \vec{\sigma}\otimes\mathds{1} + \mathds{1}\otimes\vec{b}\vec{\sigma} + \sum_{m,n=1}^{3} t_{mn} \sigma_{m}\otimes\sigma_{n}) \label{2x2bloch2}\end{equation}
Since a single qubit state corresponds to a Bloch vector, i.e. a point of a three dimensional sphere, any product state of two qubits corresponds to two distinct Bloch vectors, i.e. $t_{mn} = a_{m} b_{n}$, spanning a 6-dimensional subspace of the 15-dimensional state space (hence, all product states of two qubits can be considered as points on a 6-dimensional sphere). As soon as the two qubits are correlated -- either classically or via entanglement --, this picture does not hold anymore.\\ 

\begin{figure}[ht!]\centering\includegraphics[width=80mm]{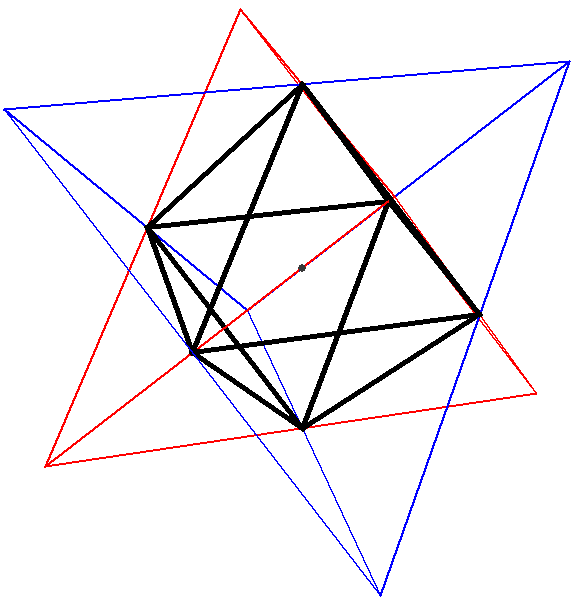}\caption[Visualisation of the maxic simplex in a system of two qubits]{Visualisation of the maxic simplex in a system of two qubits. The set of all states in the simplex form a tetrahedron (blue), so do all matrices that are positive under partial transposition (red). The intersection of these two tetrahedra is the set of PPT states, i.e. the set of separable states (black). The maximally mixed state is located in the middle of this set (gray dot), while the four Bell states are the states furthest away from it (i.e. the four vertices of the blue tetrahedron).}\label{fig_2x2ms}\end{figure}

\begin{center}\emph{The magic simplex}\end{center}
Since the whole 15-dimensional state space can hardly be visualised, often lower dimensional simplices are studied. A region of special interest is the so called magic simplex, which is the subspace spanned by four mutually orthogonal maximally entangled states (which can without loss of generality be chosen to be the four Bell states (\ref{bellstates})). Consequently, all of these states have maximally mixed subsystems, hence the local components $\vec{a}$ and $\vec{b}$ in the Bloch vector representation (\ref{2x2bloch2}) equal zero. Since the matrix $t_{mn}$ can be diagonalised, all these states are of the form
\begin{equation} \rho = \frac{1}{4}\left(\mathds{1} + \sum_{i=1}^{3} c_{i} \sigma_{i}\otimes\sigma_{i}\right) \end{equation}
where $c_{i}$ are the elements of the diagonalised matrix $t_{mn}$. More explicitly, these states can also be written as
\begin{equation} \rho = a_{1} \left|\Psi^{+}\right\rangle\left\langle\Psi^{+}\right| + a_{2} \left|\Psi^{-}\right\rangle\left\langle\Psi^{-}\right| + a_{3} \left|\Phi^{+}\right\rangle\left\langle\Phi^{+}\right| + (1-a_{1}-a_{2}-a_{3}) \left|\Phi^{-}\right\rangle\left\langle\Phi^{-}\right| \end{equation}
with $a_{i} \geq 0$ and $\sum a_{i} = 1$.\\
Since the four Bell states form the vertices of the magic simplex and are mutually equally distant from each other (in the Hilbert-Schmidt metric), the magic simplex has the shape of a tetrahedron\cite{bertl_narn_thirring}. The set of all matrices that satisfy the PPT-criterion form another tetrahedron, which however does not fully consist of states (as there are matrices that are not positive semidefinite, but become so after partial transposition). The intersection of these two tetrahedra forms the set of all separable states, which has the shape of an octahedron.\\
Thus, the magic simplex is a tetrahedron enclosing an octahedron, where the vertices of the tetrahedron are the maximally entangled states (which are the only pure states in the simplex) and the octahedron is formed by all separable states. The maximally mixed state is located in the middle of the octahedron (see Fig. \ref{fig_2x2ms}).\\

\subsubsection{Systems of two QuTrits}
Despite much effort that has been put into the understanding of the geometry of two qutrit systems, there is still rather little known about it. The existence of PPT bound entangled states (which are hard to discriminate from separable states) as well as the assumed existence of NPT bound entanglement (which is very hard to distinguish from free entangled states) and the nonexistence of analytical means to detect nonlocality pose considerable problems.\\

\begin{center}\emph{Construction of the magic simplex}\end{center}
It is possible to construct a magic simplex for $3\otimes3$ systems as well, analogously to the $2\otimes2$ case, as a mixture of a complete set of mutually orthogonal maximally entangled states. This works as follows\cite{3x3magicsimplex}:\\
One first choses any maximally entangled state $\left|\Omega_{0,0}\right\rangle$ and choses the basis such that
\begin{equation} \left|\Omega_{0,0}\right\rangle = \frac{1}{\sqrt{3}} \sum_{i=1}^{3} \left|i, i\right\rangle\end{equation}
One then defindes 8 other states
\begin{equation} \left|\Omega_{k,l}\right\rangle = W_{k,l}\otimes\mathds{1} \left|\Omega_{0,0}\right\rangle \end{equation}
via the Weyl operators $W_{k,l}$, which are defined in eq. (\ref{weyl}).\\
Since the Weyl operators are unitary, these states are maximally entangled as well and are defined such that all $\left|\Omega_{k,l}\right\rangle$ are mutually orthogonal. Now, the magic simplex $\mathcal{W}$ can be defined as the set of all states that can be written as
\begin{equation} \rho = \sum_{k,l} c_{k,l} P_{k,l} \end{equation}
where $P_{k,l} = \left|\Omega_{k,l}\right\rangle\left\langle\Omega_{k,l}\right|$ and the $c_{k,l}$ are probabilities, i.e. $c_{k,l} \geq 0$ and $\sum c_{k,l} = 1$.\\

\begin{figure}[ht!]\centering\includegraphics[width=80mm]{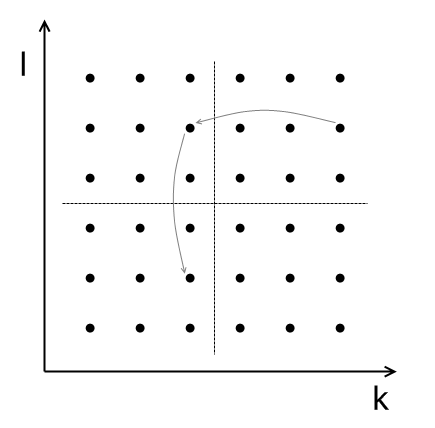}\caption[Periodic symmetry in the magic simplex of two qutrits]{The magic simplex in a system of two qutrits has a periodic symmetry in the form of a lattice, where each point corresponds to a maximally entangled state $P_{k,l}$. Each of the cells containing 9 states is fully equivalent to each other such cell, such that it suffices to study states $P_{k,l}$ with indices $k,l \in \{0,1,2\}$. This structure is commonly referred to as a phase space structure.}\label{fig_phasespace}\end{figure}

\begin{center}\emph{Geometrical properties of the magic simplex}\end{center}
Since the Weyl operators, that were used to define the vertices of $\mathcal{W}$ form a discrete group, $\mathcal{W}$ itself is highly symmetrical. Repetitive application of any $W_{k,l}\otimes\mathds{1}$ to any one of the $\left|\Omega_{k,l}\right\rangle$ always forms a set of three different states $\left|\Omega_{k,l}\right\rangle, \left|\Omega_{m,n}\right\rangle$ and $\left|\Omega_{2m-k, 2n-l}\right\rangle$. These states can be visualised as being located on a line, while all maximally entangled states form a periodic lattice (Fig. \ref{fig_phasespace}). Obviously, any two states automatically form a line, while a third state can either be on this line or not. Thus, out of all the 84 sets that contain three of the nine maximally entangled states, there are 12 that represent a line. These have several special properties, most importantly, to each of the 12 lines corresponds one of the 12 furthest outward separable states
\begin{equation} \sigma_{out} = \frac{1}{3} \sum_{(k,l) \in \mathrm{line}} P_{k,l} \end{equation}
Evidently, each of the nine maximally entangled states is equivalent to another (as long as they are considered by themselves and not in the context of specific other states). Since any two of the states automatically form a line and thus have similar geometric properties, there is also only one equivalence class for pairs of states. For triplets of states, there obviously exist two equivalence classes, one for all triplets that form a line and one for all that do not. For quadruplets of states there also exist two distinct equivalence classes, one in which three out of the four states form a line and one in which there is no line at all. For the complementary sets, the same statements hold (i.e. there are two equivalence classes of sets of five states each, two classes of sets of six states each, one class of septuplets and one class of octuplets).\\

\begin{center}\emph{Mathematical properties of the magic simplex}\end{center}
By definition, all states in $\mathcal{W}$ have maximally mixed subsystems. However, unlike in the $2\otimes2$ case, the converse statement is not true: There are states with maximally mixed subsystems, that do not belong to $\mathcal{W}$, for example the state
\[ \rho = \frac{1}{3} \left|\Psi\right\rangle\left\langle\Psi\right| + \frac{2}{3}\left|\Phi\right\rangle\left\langle\Phi\right| \]
\begin{equation} \left|\Psi\right\rangle = \left|0,0\right\rangle \end{equation}
\[ \left|\Phi\right\rangle = \frac{1}{\sqrt{2}}\left(\left|1,1\right\rangle + \left|2,2\right\rangle\right) \]
Also, the magic simplex is not unique, as it is in the $2\otimes2$ case. There exist several inequivalent sets of maximally entangled states, that result in different geometrical properties of the simplex formed by them. Due to the symmetry, choice of three different maximally entangled states that do not form a line determines the simplex.\\ \newpage

\begin{figure}[ht!]\centering\includegraphics[width=100mm]{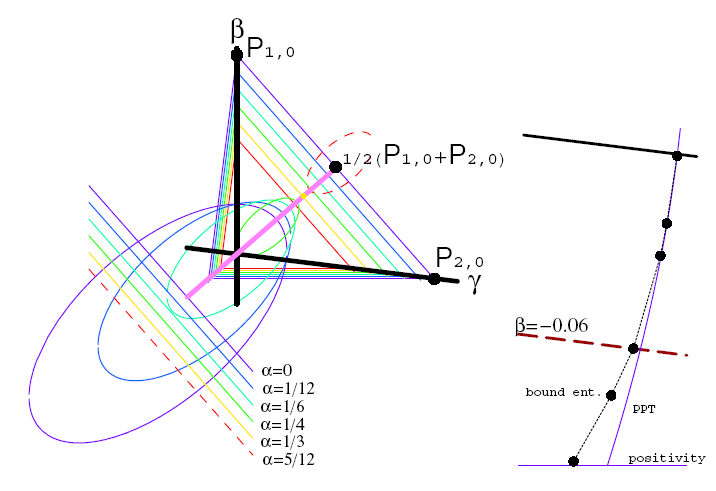}\caption[Geometry of states of the form $\rho = \frac{1-\alpha-\beta-\gamma}{9}\mathds{1} + \alpha P_{0,0} + \beta P_{1,0} + \gamma P_{2,0}$, taken from \cite{3x3geombe}]{States of the form (\ref{eq_fig4}) -- here, without loss of generality $k = l = n = 0$ and $m = 1$ -- exhibit bound entanglement only for negative parameter values. The triangles indicate areas of positivity, i.e. the set of states. PPT areas are ellipses cut by lines (or simply ellipses, for higher values of $\alpha$). For $\alpha = 0$, the small area of bound entanglement is enlarged in the right picture, where the dots correspond to points where the used entanglement witnesses are optimal\cite{3x3geombe}.}\label{fig4}\end{figure}

\begin{figure}[ht!]\centering\includegraphics[width=80mm]{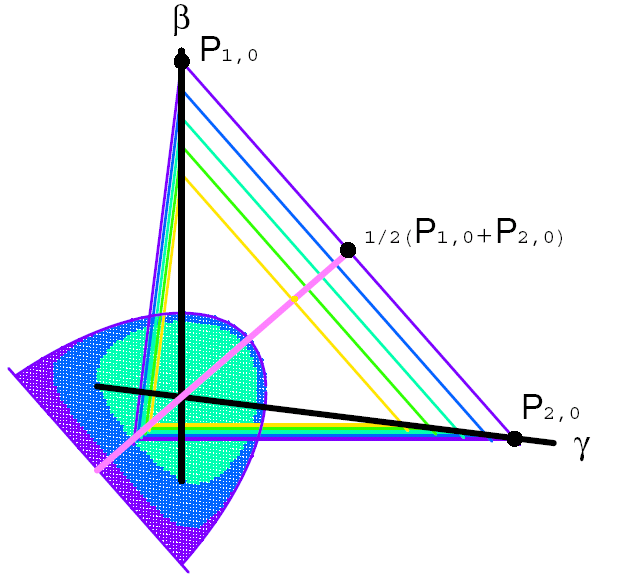}\caption[Geometry of states of the form $\rho = \frac{1-\alpha-\beta-\gamma}{9}\mathds{1} + \alpha P_{1,0} + \beta P_{2,0} + \gamma P_{1,1}$, taken from \cite{3x3geombe}]{States of the form (\ref{eq_fig6}) -- here, without loss of generality $l = n = 0$, $k = m = p = 1$ and $o = -1$ (or, equivalently, $o = 2$) -- show much more complicated structures than those of the form (\ref{eq_fig4}). The set of all states still is triangularly shaped, while the set of PPT states no more has the form of an ellipse, but a more complex one, part of which consists not of separable but of bound entangled states. The colour code is similar to Fig. \ref{fig4}\cite{3x3geombe}.}\label{fig6}\end{figure}

\begin{center}\emph{Studying families of states}\end{center}
Since $\mathcal{W}$ is 8-dimensional and it is thus not possible to graphically visualise it, further subsets of dimension two or three are of great interest. Usually, mainly mixtures of two or three maximally entangled states and the maximally mixed state are investigated, since these seem to offer most information about the geometry of the system.\\
Considering for example a mixture of three maximally entangled states on a line and the maximally mixed state\cite{3x3geombe}
\begin{equation} \rho = \frac{1-\alpha-\beta-\gamma}{9}\mathds{1} + \alpha P_{k,l} + \beta P_{m,n} + P_{2m-k,2n-l} \label{eq_fig4}\end{equation}
with $\alpha, \beta, \gamma \geq 0$ and $\alpha + \beta + \gamma \leq 1$, one finds that these states are still rather simple, since they do not contain any bound entanglement and have a very high symmetry. If however one also admits pseudomixtures (i.e. allows all values for $\alpha, \beta$ and $\gamma$ that correspond to density matrices, in particular also negative values), small regions of bound entanglement appear (see Fig. \ref{fig4}).\\
Even more complex sets of states can be found by considering states that are not located on a line
\begin{equation} \rho = \frac{1-\alpha-\beta-\gamma}{9}\mathds{1} + \alpha P_{k,l} + \beta P_{m,n} + P_{o,p} \label{eq_fig6}\end{equation}
where either $o \neq 2m-k$ or $p \neq 2n-l$ or both. Here the regions of bound entanglement are much bigger and the space shows comperatively complex structures (see Fig. \ref{fig6}).\\
In attempts to understand the geometry of mixtures of more states, mixtures of many states having only few different weights have been studied\cite{bein3x3, geomofent3x3}, revealing that bound entanglement is not at all rare (although difficult to detect) and demonstrating that even such high mixtures still show high symmetry, since the magic simplex itself does.\\

\newpage
\vspace*{7mm}
\section{Conclusion}
Ever since the field of quantum information has been started to be taken seriously, there has been great advancement especially in characterising bipartite entangled systems (while multipartite entanglement is still widely unexplored).\\
While comperatively low dimensional systems, such as systems of two qubits or one qubit and one qutrit, as well as arbitrarily dimensional pure states, hardly seem to hold any secrets anymore, systems start to behave in very complex and unanticipated ways as soon as the dimensions get higher (starting with systems of two qutrits, which already are 80-dimensional). States in these systems can show completely different features from the lower dimensional ones, the most remarkable of which is probably the phenomenon of bound entanglement.\\
Discriminating between entangled and separable states becomes difficult, because tools like the PPT criterion are not capable of answering this question definitely anymore in these cases. Although stronger tools such as entanglement witnesses are at hand, these are much more difficult to put to use, thus still making it a hard task to determine any given state's entanglement properties.\\
The question of quantifying entanglement also becomes a very tricky one in higher dimensions, as it is not quite clear, which requirements a measure for entanglement should satisfy. In fact, it seems as if a single measure does not suffice at all to completely characterise and quantise the entanglement of general states. It is also completely unknown how a set of entanglement measures would need to be organised, in order to be 'complete' in the sense of containing full information about the entanglement contained in a quantum state.\\
Several entanglement measures are already widely established and have been thoroughly studied. Some of these indeed seem to work very well even in high dimensional systems, with the downside that most of them are (at the moment) not analytically computable (or even not computable at all). The probably most important and promising entanglement measures are the entanglement of formation and the entanglement of distillation.\\ \\
The more work is put into understanding the structure of the underlying state spaces, the more it becomes appearent that geometry and symmetry play very important roles and often offer a very intuitive and vivid (although maybe sometimes misleading) picture. This allows for geometrically motivated tools, such as geometric entanglement witnesses or geometric entanglement measures to be used in a very natural and sometimes surprisingly nonmathematical way.\\ \\
While low-dimensional quantum information theory already finds realisation not only in laboratories but also is on the verge of commercial application, higher dimensional quantum information is still very young and only beginning to be understood. New effects and phenomenae are found frequently and there is no way of telling how many complex, contraintuitive and amazing discoveries there are yet to come in this field.\\ \\

\vspace{35mm}
\begin{center}\large{\textbf{Acknowledgements}}\end{center}
I would like to thank everybody without whom I would not have found my fascination for physics. In particular this primarily means my father, who showed me the stunning beauty of nature and science throughout my childhood and youth, and Helmut Linhart, a physics teacher at my former school who introduced me to the amazement of theoretical physics and quantum mechanics and was probably one of the main reasons I started studying physics five years ago.\\
I am also very grateful for all my friends and colleagues who supported me on this way -- be it privately or professionally.\\

\newpage
\vspace*{30mm}
\listoffigures
\newpage
\bibliography{Diplomarbeit}
\bibliographystyle{unsrt}

\end{document}